\documentclass[onecolumn,journal,noadjust]{IEEEtran}
\usepackage{cite}

\ifCLASSINFOpdf
\else
\fi
\usepackage{amssymb}
\usepackage{dsfont}
\usepackage{float}
\usepackage{graphicx}
\usepackage{xcolor}
\newtheorem{lemma}{Lemma}
\newtheorem{fact}{Fact}
\newtheorem{definition}{Definition}
\newtheorem{proposition}{Proposition}
\newtheorem{theorem}{Theorem}
\newtheorem{remark}{Remark}

\usepackage [autostyle, english = american]{csquotes}
\MakeOuterQuote{"}

\usepackage{amsmath}
\begin{document}
%
\title{Lossy Compression with Universal Distortion}
%
%
%

\author{Adeel Mahmood and Aaron B. Wagner\\
School of Electrical and Computer Engineering, Cornell University}

\maketitle

\begin{abstract}
We consider a novel variant of $d$-semifaithful lossy coding in which the distortion measure is revealed only to the encoder and only at run-time, as well as an extension of it in which the distortion constraint $d$ is also revealed at run-time. Two forms of rate redundancy are used to analyze the performance, and achievability results of both a pointwise and minimax nature are demonstrated. The first coding scheme uses ideas from VC dimension and growth functions, the second uses appropriate quantization of the space of distortion measures, and the third relies on a random coding argument.   
\end{abstract}


%
\IEEEpeerreviewmaketitle

\section{Introduction}
%
%
%
%
Lossless coding is the mapping of raw data to a binary representation such that the original data can be exactly recovered from the binary representation. For mathematical analysis, the raw data is treated as a randomly generated \textit{source sequence} and the corresponding binary representation is in the form of a binary string. In this paper, we will focus on discrete and memoryless sources, i.e., each source symbol in the sequence is independent and identically distributed and takes values on a finite alphabet. A lossless \textit{encoder} carries out the source-to-binary mapping while a decoder performs the inverse mapping. Together, the encoder and decoder pair specify a coding scheme. The performance of a lossless coding scheme is usually\footnote{Other performance metrics such as the probabilistic $\epsilon$-length \cite{kosut1} are also used.} measured by the expected length of the binary string per source symbol (or simply the expected rate), where the expectation is with respect to (w.r.t.) the source probability distribution. Shannon entropy of the source probability distribution characterizes the minimum (asymptotically) achievable expected rate\footnote{For prefix-free lossless codes \cite[Theorem 5.3.1]{thomas_cov}}. A precise performance metric is, therefore, the difference between the expected rate and Shannon entropy. This is called the \emph{rate redundancy}.

In lossy coding, the original source sequence is not recovered exactly and is instead approximated by what is called a reconstruction sequence. The rate redundancy in lossy coding is defined similarly, except that the rate-distortion function \cite{berger1} now plays the role of Shannon entropy. In this paper, we will focus on a generalization of \emph{$d$-semifaithful coding}~\cite{Ornstein:Semifaithful}, a form of lossy compression in which the decoder outputs a reconstruction sequence that is within distortion $d$ of the original source sequence with probability one. Distortion is measured by a single-letter distortion measure which we will denote by $\rho$. Denoting the length of the source sequence, also called the block length, by $n$, past work has analyzed the rate of convergence of the average expected codeword length to the rate-distortion function as a function of $n$.  \cite[Theorem 5]{yang1} established an achievable rate redundancy of $\ln n/n + o\left( \ln n/n \right)$ while \cite[Theorem 4]{yang1} established a converse of $1/2\ln n/n + o \left(\ln n/n \right)$. 

Universal coding schemes are of interest when the source probability distribution $p$ is unknown. A coding scheme is said to be \emph{universal} over a class of source distributions if the rate redundancy converges to zero for every source in that class. If the convergence is pointwise, then we say the coding scheme is \emph{weakly universal}. If the convergence is uniform (or minimax), then the coding scheme is \emph{strongly universal}. These two notions of universality originated in the universal noiseless coding literature \cite{1055092}.  Let $J$ and $K$ be the sizes of source and reconstruction alphabets, respectively. Yu and Speed \cite[Theorem 2]{yu1} established an achievable weakly universal convergence rate of 
\begin{align}
    (KJ + J + 4)\frac{\log n}{n} + O(n^{-1}) 
    \label{yuka}
\end{align}
for the rate redundancy of universal $d$-semifaithul codes for a class of source distributions $p$ satisfying some regularity conditions. On the other hand, one can also consider a modified rate redundancy, replacing the rate-distortion function with Shannon entropy of the probability distribution of reconstruction sequences, minimized over all $d$-semifaithful codes, see \cite{508836aa}, \cite{silva1}. This form of rate redundancy essentially considers the difference between the expected rate of a given universal code and the expected rate of an optimal $n$th order code. Throughout the paper, the rate redundancy w.r.t. the rate-distortion function will be referred to as simply the rate redundancy while the latter formulation will be called the \textit{operational} rate redundancy. With the operational rate redundancy as the metric, one can establish (e.g., \cite[Lemma 4]{silva1}) an achievable strongly universal convergence rate of  
\begin{align}
    (J-1) \frac{\log n}{ n} + O(n^{-1}). \label{eq:silva1}
\end{align}
In both results $(\ref{yuka})$ and $(\ref{eq:silva1})$, the distortion measure $\rho$ is fixed and known to both the encoder and decoder. A novel variation of (universal) $d$-semifaithful coding (and lossy coding in general) we consider is 
that in which the distortion measure is revealed to the encoder alone, and only when it receives the source sequence $x^n$ to compress. We call this the \textit{universal distortion} problem. Traditional $d$-semifaithful coding framework can be roughly represented by  
\begin{align}
\begin{split}
    &\textbf{encoder} : x^n \longmapsto \textbf{binary string} \\
    &\textbf{decoder} : \textbf{binary string} \longmapsto y^n
    \end{split} \label{cartoon_code}
\end{align}
where $x^n$ is the given source sequence to be compressed and $y^n$ is the reconstruction sequence satisfying the distortion constraint with respect to $\rho$. Here the distortion measure $\rho$ is fixed \emph{a priori}. On the other hand, universal distortion $d$-semifaithful coding can be represented by
\begin{align}
\begin{split}
    &\textbf{encoder} : \left( x^n, \rho \right) \longmapsto \textbf{binary string}\\
    &\textbf{decoder} : \textbf{binary string} \longmapsto y^n
    \end{split} \label{unknown_cartoon_code}
\end{align}
We elaborate the distinction between $(\ref{cartoon_code})$ and $(\ref{unknown_cartoon_code})$ in terms of the codebook underlying the encoder and decoder pair. The task of designing a coding scheme is simplified by sharing a codebook of indexed reconstruction sequences between the encoder and decoder. In this case, the encoder transmits the index (as a binary string) of a codeword which gives smaller than $d$ distortion with the given source sequence. In traditional $d$-semifaithful coding, the codebook is optimally designed to minimize the average rate and keep distortion less than $d$ with respect to one fixed distortion measure. In the universal distortion formulation, one codebook must be rich enough to cover all source sequences with less than $d$ distortion with respect to a variety of distortion measures. An extension of this framework, which we will call the \textit{generalized} universal distortion problem, is when the distortion constraint $d$ is itself a run-time input to the encoder alone:
\begin{align}
\begin{split}
    &\textbf{encoder} : \left( x^n, \rho, d \right) \longmapsto \textbf{binary string}\\
    &\textbf{decoder} : \textbf{binary string} \longmapsto y^n
    \end{split} \label{unknown_cartoon_code_2}
\end{align}

A natural approach to the universal distortion problem is for the encoder to report a quantized version of the distortion measure to the decoder and then proceed as if the communicated distortion measure was in effect. For the universal distortion framework, we show that this simple approach (with some post-correction modification) yields a strongly universal (or minimax) achievability result with respect to the operational rate redundancy (see Theorem~\ref{specthm}). The quantization approach only works for uniformly bounded distortion measures, however. For the generalized universal distortion framework, we replace the quantization approach with one based on ideas from VC dimension theory, giving a strongly universal achievability result with respect to the operational rate redundancy (see Theorem~\ref{silvaext}), where universality now includes all unbounded distortion measures and distortion levels. Returning to the traditional rate redundancy with respect to the rate-distortion function, we use a random coding approach to give a weakly universal achievability result for the universal distortion framework (see Theorem \ref{thmyuext}). All three results have a $O(\ln n / n)$ convergence rate which is the optimal order of convergence for traditional lossy source coding \cite{yang1}; in particular, the achievability result of Theorem \ref{thmyuext} is within $\ln n / n$ of the known converse bound \cite[Theorem 1]{yang3} for traditional universal $d$-semifaithful codes and in fact, matches the best known achievability result \cite[Theorem 2]{yang3} for traditional universal $d$-semifaithful codes, while itself being a universal distortion $d$-semifaithful code. 

Subsequent to the initial version of this work \cite{https://doi.org/10.48550/arxiv.2110.07022}, Merhav \cite{https://doi.org/10.48550/arxiv.2203.03305} has provided pointwise achievability and converse results in the universal distortion framework. His rate redundancy results are with respect to the empirical rate-distortion function\footnote{The empirical rate-distortion function is equal to the rate-distortion function evaluated at the empirical distribution of a given realization of a source sequence, as opposed to the underlying source distribution itself. } and use a different random coding approach. In his context, pointwise means that the convergence rate is not uniform but depends on the source sequence through its type (and the distortion measure). As described above, our work focuses on expected rate redundancy so pointwise in our paper means for each unknown source $p$ and distortion measure. Both notions of pointwise should be considered weakly universal. Other points of comparison with \cite{https://doi.org/10.48550/arxiv.2203.03305} as well as with Yang and Zhang's earlier work (\cite{yang2}, \cite{yang3}) in traditional universal lossy coding will be laid out in the subsequent presentation of our main results.

One practical motivation for the universal distortion setup comes from the observation that compression systems
are typically asked to meet the needs of a variety of end-users who may have
discordant notions of distortion. In the context of images, for some users,
a decoder that includes artificial high-frequency components in order to
make the reconstructed image more pleasing is preferable to one that 
simply outputs blurry images, even though the high-frequency components
might not match the original image \cite{NEURIPS2018_801fd8c2}. For other end-users, the opposite will
be true. Specifically, image compression methods based on deep neural networks, which learn to synthesize local image content, can lead to large distortions with respect to traditional distortion metrics such as peak signal-to-noise ratio but perform much better with distortion metrics based on perceptual transforms \cite{balle1}. One would like to design codes that respect the distortion
constraints of the particular users using the system, which might only be 
known at run-time. In a similar vein, image compression methods based on saliency maps \cite{5223506} can be viewed within the (generalized) universal distortion framework; an image to be compressed is divided into different subblocks based on relative importance and each subblock is compressed with a different distortion.   

A separate motivation comes from nonlinear transform 
coding~\cite{balle1, balle2, goyal1}. Suppose a source $x^n$
is first mapped to a set of tranform coefficients $z^k$ via 
the analysis transform $g_a(\cdot)$,
\begin{equation}
    z^k = g_a(x^n).
\end{equation}
The transform coefficents are then quantized using some quantizer
$Q(\cdot)$,
\begin{equation}
    \hat{z}^k = Q(z^k),
\end{equation}
where the range space of $Q(\cdot)$ is discrete and heavily
constrained. A synthesis transform $g_s(\cdot)$ is then
used to create the reconstruction $y^n$:
\begin{equation}
    y^n = g_s(\hat{z}^k).
\end{equation}
In linear transform coding, the $g_a(\cdot)$ and $g_s(\cdot)$
transforms are typically isometric with respect to $L^2$
distance. Thus they are mean-squared error (MSE) preserving
and $Q(z^k)$ should map $z^k$ to the closest quantization
$\hat{z}^k$ in $L^2$ distance. 

Recently, however, promising results have been obtained
via nonlinear transform coding, specifically those obtained
via stochastic training of artificial neural 
networks~(e.g., \cite{balle1, balle2, 
Balle:Divisive, 
Balle:Hyperprior,
Agustsson:Extreme,
Agustsson:SofttoHard, Theis:LossyAuto,
Balle:NTC,Toderici:Recurrent:CVPR,
Johnston:Priming, Li:RevAE, Santurkar:Generative, Gregor:Towards}).
Such learned, nonlinear transforms are not guaranteed
to be distance-preserving, however. Thus mapping $z^k$
to the nearest quantization point is not equivalent to
finding the $\hat{z}^k$ that minimizes
\begin{equation}
    \label{eq:nonlineardist}
    \rho(x^n,g_s(\hat{z}^k)).
\end{equation}
In principle, the quantizer $Q(\cdot)$ could map a given
$z^k$ to the $\hat{z}^k$ that minimizes 
(\ref{eq:nonlineardist}); in practice, this is expensive.
An alternative is to consider a quadratic approximation of
(\ref{eq:nonlineardist}) about $\hat{z}^k = z^k$:
\begin{align}
    \rho(x^n,g_s(\hat{z}^k)) &\approx
    \rho(x^n,g_s(z^k)) + \nabla_{z^k} \rho(x^n,g_s(z^k))^T(\hat{z}^k - z^k)
          + \mbox{ } \notag\\
          &\,\,\,\,\,\,\,\,\,\,\,\,\,\, \frac{1}{2} (\hat{z}^k - z^k)^T \nabla^2_{z^k} \rho(x^n,g_s(z^k))
              (\hat{z}^k - z^k),
         \label{eq:nonlinearexpansion}
\end{align}
where $\nabla_{z^k} \rho (x^n,g_s(z^k))$ 
and $\nabla^2_{z^k} \rho (x^n,g_s(z^k))$ denote the
gradient and Hessian, respectively.
Note that the first term on the right-hand side of 
(\ref{eq:nonlinearexpansion}) does not depend on $\hat{z}^k$.
Thus minimizing
(\ref{eq:nonlinearexpansion}) is tantamount to minimizing
\begin{align}
     & \nabla_{z^k} \rho(x^n,g_s(z^k))^T(\hat{z}^k - z^k)
          + \mbox{ } \nonumber \\
    & \phantom{\nabla} \frac{1}{2} (\hat{z}^k - z^k)^T \nabla^2_{z^k} 
        \rho(x^n,g_s(z^k)) (\hat{z}^k - z^k)
    \label{eq:nonlinearquadratic}
\end{align}
over $\hat{z}^k$.
We arrive at the problem studied in this paper, in
which we seek to quantize a given source realization 
$z^k$ according to a distortion measure
that is not known until $z^k$ itself is known.

For transforms that are trained \emph{end-to-end},
there is evidence that the Jacobian of $g_s(z^k)$,
when viewed as a $k$-by-$n$ matrix, has orthonormal
rows with high probability~\cite[supp. mat.]{Balle:NTC}.
If the gradient $\nabla_{z^k} \rho(x^n,g_s(z^k))$ is
also zero, then the first term in (\ref{eq:nonlinearquadratic})
vanishes and the Hessian is proportional to the identity
matrix, eliminating the need for distortion universality. A number
of nonlinear transforms have been proposed for compression
that are not trained in this fashion, however~\cite{Toderici:Recurrent:CVPR,
Johnston:Priming, Li:RevAE, Santurkar:Generative, Gregor:Towards}.
Even for those that are, employing a quantizer that minimizes
the objective in (\ref{eq:nonlinearquadratic}) could allow for 
reduced capacity
in the neural networks comprising the analysis and synthesis
transforms, with a concomitant reduction in training requirements.
Application to nonlinear transform coding was the original 
motivation for this work.

\section{Preliminaries}

Let $A$ and $B$ denote finite source and reconstruction alphabets, respectively. 
Without loss of generality, we can let $A = \{1, 2, ..., J \}$ and $B = \{1,2, ..., K \}$. $\mathcal{P}(A)$ denotes the set of all probability distributions on $A$. $\mathcal{P}(A|B)$ denotes the set of all conditional distributions. In this paper, $\,\ln\,$ represents log to the base $e$, $\,\log\,$ represents log to the base $2$ and $\exp(x)$ is equal to $e$ to the power of $x$. Unless otherwise stated, all information theoretic quantities will be measured in nats. For $p \in \mathcal{P}(A)$, $H(p)$ denotes the Shannon entropy. For $p \in \mathcal{P}(A)$ and $W \in \mathcal{P}(B|A)$, $H(W|p)$ denotes the conditional entropy and $I(p, W) = I(X;Y)$ denotes the mutual information where $(X,Y)$ have the joint distribution given by $p \times W$. 

For $p_1 \in \mathcal{P}(A)$ and $p_2 \in \mathcal{P}(A)$,
$D(p_1 || p_2)$ denotes the relative entropy between the two probability distributions. For any vector $v \in \mathbb{R}^m$, $||v||_1$ and $||v||_2$ will denote the $l^1$ and $l^2$ norms of $v$, respectively. For any two $m$-dimensional vectors $u = (u_1, ..., u_m)$ and $v = (v_1, ..., v_m)$, $\delta(v, u) \triangleq 1/2 ||v - u||_1 $ will denote their total variation distance. We will frequently view probability distributions $p \in \mathcal{P}(A)$  as $J$-dimensional vectors. Finally, for any matrix $M \in \mathbb{R}^{n \times m}$, $||M||_F$ will denote the Frobenius norm of $M$.  

For a given sequence $x^n \in A^n$, the $n$-type $t = t(x^n)$ of $x^n$ is defined as
\begin{align*}
    t(j) &= \frac{1}{n} \sum_{i=1}^n \mathds{1}(x_i = j)
\end{align*}
for all $j \in A$, where $\mathds{1}(\cdot)$ is the indicator function. $\mathcal{P}_n(A)$ denotes the set of all $n$-types on $A$. For a pair of sequences $x^n \in A^n$ and $y^n \in B^n$, the joint $n$-type $s$ is defined as 
\begin{align*}
    s(j, k) &= \frac{1}{n} \sum_{i=1}^n \mathds{1} \left( x_i = j, y_i = k \right)
\end{align*}
for all $j \in A$ and $k \in B$. $\mathcal{P}_n(A \times B)$ denotes the set of all joint $n$-types on $A \times B$. For two sequences $x^n$ and $y^n$ with $n$-types $t_x = t(x^n)$ and $t_y = t(y^n)$, the joint $n$-type $s $ can also be written as
\begin{align*}
    s(j, k) &= t_x(j) W_y(k|j) = t_y(k) W_x(j|k),
\end{align*}
where $W_y$ is called a conditional type of $y^n$ given $x^n$, and $W_x$ is called a conditional type of $x^n$ given $y^n$. From \cite[Lemma 2.2]{korner1}, we have 
\begin{align}
\begin{split}
    |\mathcal{P}_n(A)| &\leq (n+1)^{J-1}  \text{ and} \\
    |\mathcal{P}_n(A \times B)| &\leq (n+1)^{JK - 1}.
\end{split}
\label{ghba}
\end{align}
For a given type $t \in \mathcal{P}_n(A)$, $T_A^n(t)$ is called the type class where
\begin{align}
    T_A^n(t) &\triangleq \{x^n \in A^n: t(x^n) = t  \}.\notag 
\end{align}
For any given $p \in \mathcal{P}(A)$ or $p \in \mathcal{P}(B)$, $p^n$ will denote the $n$-fold product distribution induced by $p$. Let $X^n$ be an independent and identically distributed source. Let $p \in \mathcal{P}(A)$ be the generic probability distribution of the source so that $X^n$ is distributed according to $p^n$. The probability that $X^n$ is of type $t$ is given by \cite[Lemma 2.6]{korner1}
\begin{align}
    \mathbb{P}_p \left( X^n \in T^n_A(t) \right) &= p^n \left( T^n_A(t)\right) \leq \exp \left(  - n D(t||p) \right). \label{opqq}
\end{align}

For a given source distribution $p$, it suffices to focus only on sequence types $t$ satisfying $||t - p||_2 \leq a \sqrt{\ln n/n}$, where $a^2 \geq 2 + 2J$. Source sequence types farther away from the source distribution $p$ have negligible probability for large $n$ as quantified by the following lemma. 
\begin{lemma}
If $a \in \mathbb{R}_{\geq 0}$ satisfies $a^2 \geq 2 + 2J$, then for all $p \in \mathcal{P}(A)$ and all $n \in \mathbb{N}$ , we have 
\begin{align*}
    \sum_{t:||t - p||_2 > a \sqrt{\ln n / n}} p^n(T^n_A(t)) \leq \frac{e^{J-1}}{n^2}.
\end{align*}
\label{lemmatypes}
\end{lemma}
\begin{IEEEproof}[Proof]
For any type $t$ satisfying $||t - p||_2 > a \sqrt{\ln n/n}$, we have $\delta(t,p) \geq \frac{1}{2} ||t - p||_2 > \frac{1}{2} a \sqrt{\ln n / n}$ where $\delta(t,p)$ is the total variation distance and the inequality follows by the fact that the Euclidean distance is upper bounded by the $l^1$ norm. By Pinsker's inequality \cite[3.18]{korner1}, we then have 
\begin{align*}
    D(t||p) &\geq 2 \delta^2(t,p) \geq   \frac{a^2 \ln n}{2n}.
\end{align*}
If $a \geq \sqrt{2 + 2J}$, we have 
\begin{align}
        \sum_{t:||t - p||_2 > a \sqrt{\ln n/n}}p^n\left( T^n_A(t)\right) &\leq \sum_{t:||t - p||_2 > a \sqrt{\ln n/n}} e^{-n D(t||p)} \notag \\
        &\leq (n+1)^{J-1} e^{-a^2 \ln n/2} \notag \\
        &\leq e^{J-1} e^{(J-1)\ln n} e^{-a^2  \ln n/2} \notag \\
        &\leq \frac{e^{J-1}}{n^{2}}.
        \notag
\end{align}
\end{IEEEproof}

Let $\rho : A \times B \to [0, \infty )$ be a single-letter distortion measure and $\rho_n(x^n, y^n)$ be its $n$-fold extension defined as 
\begin{align}
    \rho_n(x^n, y^n) = \frac{1}{n} \sum_{i = 1}^n \rho(x_i, y_i),     \label{dist_n_fold}
\end{align}
where $x^n \in A^n$, $y^n \in B^n$. For convenience, we also define 
\begin{align}
    \rho(p, W_{B|A}) &= \sum_{j \in A, k \in B} p(j) W_{B|A}(k|j) \rho(j, k), \label{expdist}
\end{align}
which is equal to the expected distortion $\mathbb{E}[\rho(X,Y)]$ where $(X,Y)$ have the joint distribution $p \times W_{B|A}$ for some $p \in \mathcal{P}(A)$ and conditional distribution $W_{B|A} \in \mathcal{P}(B|A)$. We will frequently view distortion measures as $J \times K$ matrices, i.e., $\rho \in \mathbb{R}^{J \times K}$.  

Let $\mathcal{D}$ denote the space of all distortion measures and let $\mathcal{D}^{\rho_{\max}} \subset \mathcal{D}$ be the space of uniformly bounded distortion measures, i.e., all $\rho \in \mathcal{D}^{\rho_{\max}}$ satisfy $\rho(\cdot \, , \cdot) \leq \rho_{\max}$ for some fixed $\rho_{\max} > 0$. The results of Theorems \ref{specthm} and \ref{thmyuext}  hold only for distortion measures in $\mathcal{D}^{\rho_{\max}}$. Theorem \ref{silvaext}, on the other hand, is valid for all distortion measures in $\mathcal{D}$. Furthermore, we will use the customary assumption \cite{silva1}, \cite{kontoyiannis1}, \cite{yang2}:
\begin{align}
    \max_{j \in A} \min_{k \in B} \rho(j,k) = 0 \,\,\,\,\,\,\,\,\,\,\,\,\,\,\,\,\,\, \text{for all } \rho \in \mathcal{D}. \label{dist_assump}  
\end{align}
When the source distribution and the distortion measure are fixed, (\ref{dist_assump}) is without loss of generality~\cite[p.~26]{berger1}. Here, it is tantamount to having $d$ represent the allowable excess expected distortion above the minimum possible for the given source distribution and distortion measure. For universal distortion, this is preferable to having $d$ represent a constraint on the absolute expected distortion: a given $d$ will be below the minimum achievable expected distortion for some distortion measures, for instance.

For a given $\rho \in \mathcal{D}$, $p \in \mathcal{P}(A)$ and $d > 0$, the rate-distortion function $R(p, d, \rho)$ is defined as \cite[Theorem 10.2.1]{thomas_cov}
\begin{align}
    &R(p,d,\rho) \notag\\
    &\triangleq \min_{W_{B|A} \in \mathcal{W}_{d,\rho}} I(p, W_{B|A}) \label{rdfunc} \\
    &= \min_{W_{B|A} \in \mathcal{W}_{d,\rho}} \sum_{j, k} p(j) W_{B|A}(k|j) \ln \left(\frac{W_{B|A}(k|j)}{W(k)} \right), \notag \\
    &\text{where } W(k) = \sum_{j \in A} p(j) W_{B|A}(k|j) \text{ and }  \label{rdfunc2} \\
    \mathcal{W}_{d,\rho} &= \left \{ W_{B|A} : \sum_{j, k} p(j) W_{B|A}(k|j) \rho(j,k) \leq d \right \}. \label{rdfunc3}
\end{align}
For any given $p$ and $\rho$, $R(p,d,\rho)$ is nonincreasing, convex and differentiable  everywhere as a function of $d$ except possibly at $d = \min_{k \in B} \sum_{j \in A} p(j) \rho(j, k)$ \cite{berger1}, \cite[Exercise 8.6]{korner1}, \cite[Lemma 10.4.1]{thomas_cov}. In particular, for $0 < d < \min_{k \in B} \sum_{j \in A} p(j) \rho(j, k)$, $R(p,d,\rho)$ is strictly decreasing in $d$. 
The function's dependence on $p$ for given $d$ and $\rho$ is complex~\cite{Ahlswede:RD}.
In particular, it is not concave in general. For a given $d > 0$ and $\rho \in \mathcal{D}$, we call $R(T,d,\rho)$ the \emph{plug-in} estimator for $R(p,d,\rho)$, where $T = t(X^n)$ is the type of an i.i.d. sequence $X^n \sim p^n$.  The expected value of the estimator is given by
\begin{align*}
    \mathbb{E}_p \left [ R(T,d,\rho) \right ] &= \sum_{t \in \mathcal{P}_n(A)} p^n(T^n_A(t)) R(t,d,\rho).
\end{align*}
Harrison and Kontoyiannis \cite{harrison1} gave sufficient conditions for the consistency of the plug-in estimator. In particular, it follows from \cite[Corollary 1]{harrison1} that under the assumption in $(\ref{dist_assump})$, $R(T,d,\rho)$ is a consistent estimator for $R(p,d,\rho)$.

Throughout the paper, we will have $W^*_{B|A}$ denote an optimal transition probability matrix which achieves the minimum in (\ref{rdfunc})-(\ref{rdfunc3}). Note that $W^*_{B|A}$ is not necessarily unique, and $W^*_{B|A}$ depends on $p$, $d$ and $\rho$; when necessary, we will indicate this dependence by writing $W^*_{B|A}[p,d,\rho]$. We will use $Q^{p,d,\rho}$ to denote the corresponding optimal output distribution on $B$ associated with the optimal channel $W^*_{B|A}$, i.e., 
\begin{align*}
Q^{p,d,\rho}(k) &\triangleq \sum_{j \in A} p(j) W^*_{B|A}(k|j)     
\end{align*}
for all $k \in B$. The next lemma shows that if $W^*_{B|A}$ is unique for a particular $(p,d,\rho)$ triple, then it is continuous at this point. 
\begin{lemma}
    Fix any $p \in \mathcal{P}(A)$, $\rho \in \mathcal{D}$ and $d$ satisfying $0 < d < \min_{k \in B} \sum_{j \in A} p(j) \rho(j, k)$. Let $W^*_{B|A}$ be an optimal transition probability matrix corresponding to $(p,d,\rho)$ which achieves the minimum in $(\ref{rdfunc})$. If $W^*_{B|A}$ is the unique minimizer at $(p,d,\rho)$, then for every $\epsilon > 0$, there exists a $\delta > 0$ such that for every $(p', d',\rho') \in \mathcal{N}(p, d, \rho)$, where 
\begin{align*}
    \mathcal{N}(p, d, \rho) \triangleq &\left \{ (p',d',\rho') : ||p' - p||_2 \leq \delta, \right.\\
    &\,\,\,\,\,\,\,\,\,\,\,\,\,\,\,\,\,\,\,\,\,\,\,\,\,\, \left . |d' - d| \leq \delta \,\, \text{and}  \,\, ||\rho'-\rho||_F \leq \delta  \right \},
\end{align*}
we have \begin{align*}
    ||W^*_{B|A}[p,d,\rho] - W^*_{B|A}[p', d', \rho'] ||_F \leq \epsilon.
\end{align*}
\label{contlemma}
\end{lemma}
\begin{remark}
    Note that $W^*_{B|A}$ is not required to be unique for all points in the neighborhood
      $\mathcal{N}(p, d, \rho)$.
\end{remark}
\textit{Proof:} The proof of Lemma \ref{contlemma} is given in Appendix \ref{contlemmaproof}. 
\noindent 
\\

Previous works on lossy coding \cite{linder1}, \cite{yu1}, \cite{yang1} and \cite{yang2} have primarily considered two kinds of block codes:
\begin{itemize}
    \item fixed rate codes 
    \item $d$-semifaithful codes
\end{itemize}
As mentioned before, we will focus on the latter. An $n$th order $d$-semifaithful block code is defined by a triplet $C_n = (\phi_n, f_n, g_n)$ such that
\begin{align}
\begin{split}
    \phi_n & : A^n \to B_{\phi_n} \subset B^n\\
    f_n & : B_{\phi_n} \to \mathcal{B}^*\\
    g_n & : \mathcal{B}^* \to B_{\phi_n}
\end{split} \label{dsemi}
\end{align}
where 
\begin{itemize}
    \item $\mathcal{B}^*$ is a set of binary strings,
    \item $(f_n, g_n)$ is a prefix-free binary encoder and decoder pair,
    \item $B_{\phi_n}$ is the \text{codebook}, and
    \item $\phi_n$ is a $d$-quantizer, i.e., for all $x^n \in A^n$, we have $\rho_n(x^n, \phi_n(x^n)) \leq d.$
\end{itemize}
This formulation has been employed before~\cite{yang1,yang3}.
It should be distinguished from the definition of a $d$-semifaithful
code as a pair $(f_n',g_n')$ such that
\begin{align}
\begin{split}
    f'_n & : A^n \to  \mathcal{B}^* \\
    g'_n & : \mathcal{B}^* \to B^n,
\end{split} \label{dsemivar}
\end{align}
where 
\begin{itemize}
    \item $(f'_n, g'_n)$ is a prefix-free binary encoder and decoder pair, and
    \item for all $x^n \in A^n$, we have $\rho_n(x^n, g_n'(f_n'(x^n))) \leq d.$
\end{itemize}
Compared to (\ref{dsemivar}), the formulation in (\ref{dsemi}) incurs
a loss of generality in that it prohibits the binary encoder $f_n$ from sending
control information obtained from the input to the quantizer $\phi_n$ (but not
revealed by the codeword), such as the type of the source sequence
or a flag used to toggle between different modes of compression.
On the other hand, the structure in (\ref{dsemi}) is without loss
of optimality in that any $d$-semifaithful pair in (\ref{dsemivar})
can be reduced to a $d$-semifaithful triple in (\ref{dsemi}) 
with a rate that is only lower. Given $(f'_n,g'_n)$ in $(\ref{dsemivar})$, let
\begin{align*}
    B_{\phi_n} = \{g'_n(f_n'(x^n)) :  x^n \in A^n\}.
\end{align*}
Then define
\begin{align}
    \nonumber
    \phi_n(x^n) & = g_n'(f_n'(x^n)) \\
    \nonumber
    g_n(\cdot) & = g_n'(\cdot) \\
    \label{eq:reducedrate}
    f_n(y^n) & = \operatorname*{arg\,min}_{b \in \mathcal{B}^*: g_n'(b) = y^n}
    \ell(b) \quad \text{for} \ y^n \in B_{\phi_n}.
\end{align}
From (\ref{eq:reducedrate}) we have
\begin{equation}
    \ell(f_n(\phi_n(x^n))) \le \ell(f_n'(x^n))
\end{equation}
for all $x^n$.
We shall adopt the formulation in (\ref{dsemi}), but we shall
also allow the encoder to send control information when it is 
convenient to do so, with the understanding that the above reduction
is ultimately performed. An analogous convention will prevail
for the modified formulations of $d$-semifaithful codes given
later.

%

The performance of a $d$-semifaithful code $C_n$ can be measured by the rate redundancy $\mathcal{R}_n(C_n, p,\rho)$ defined as
\begin{align}
  \mathcal{R}_n(C_n,p,\rho) \triangleq \frac{1}{n} \mathbb{E} \left [l\left( f_n\left(\phi_n(X^n)\right) \right) \ln 2 \right ] - R(p,d,\rho), 
  \label{rate_redunda}
\end{align}
where $\mathbb{E}\left [ l(f_n(\phi_n(X^n))) \right ] $ is the expected length of the binary string $f_n(\phi_n(X^n))$, the expectation being with respect to the product distribution $p^n$ and the factor of $\ln 2$ is because we measure coding rate in nats. Note that $\mathcal{R}_n(C_n,p,\rho)$ is nonnegative for all $d$-semifaithful codes $C_n$ \cite[Secs. 5.4 and 10.4]{thomas_cov}.  

Alternatively, note that the expected length $\mathbb{E}\left [ l(f_n(\phi_n(X^n))) \right]$ of a particular $d$-semifaithful code $(\phi_n, f_n, g_n)$ is lower bounded by the Shannon entropy of the probability distribution of $\phi_n(X^n)$, where $X^n \sim p^n$ \cite[Theorem 5.3.1]{thomas_cov}. This is because the binary encoder losslessly encodes the output $\phi_n(X^n)$ of the $d$-quantizer. For a given source $p$ and $d$-quantizer $\phi_n$, the distribution of $\phi_n(X^n)$ is defined as 
\begin{align}
    \nu_{p^n, \phi_n}(y^n) &= \sum_{x^n \in A^n} p^n(x^n) \mathds{1} \left( \phi_n(x^n) = y^n \right) \label{nudef}
\end{align}
for all $y^n \in B_{\phi_n}$. Hence, an operational rate redundancy, which was considered in \cite{508836aa}, \cite{silva1}, can be defined as 
\begin{align}
    &\overline{\mathcal{R}}_n(C_n,p,\rho) \notag \\
    &\triangleq \frac{1}{n} \mathbb{E} \left [l\left( f_n\left(\phi_n(X^n)\right) \right) \ln 2 \right ] - \inf_{\phi_n \in \mathcal{Q}_{d, \rho}} \frac{H(\nu_{p^n, \phi_n})}{n}, \label{modirateredun}
\end{align}
where $\mathcal{Q}_{d,\rho}$ is the set of all possible $d$-quantizers with respect to distortion measure $\rho$. The performance metric in $(\ref{modirateredun})$ is of an operational nature; it is essentially (with a discrepancy of at most $1/n$) the difference between the expected rate of a code $C_n$ and the minimum possible expected rate of any $n$th order code, which we will call $\mathcal{M}^*(n, p, \rho)$. We can write the rate-redundancy $\mathcal{R}_n(C_n, p, \rho)$ as 
\begin{align}
\begin{split}
    &\mathcal{R}_n(C_n, p, \rho)\\
    &\approx \frac{1}{n} \mathbb{E} \left [l\left( f_n\left(\phi_n(X^n)\right) \right) \ln 2 \right ] - \mathcal{M}^*(n, p, \rho) + \mbox{}\\
    & \,\,\,\,\,\,\,\,\,\,\,\,\,\,\,\,\,\,\,\,\,\,\,\,\,\,\,\,\,\,\,\,\, \mathcal{M}^*(n, p, \rho) - R(p,d,\rho). 
\end{split}
\label{helloremiwolf}
\end{align}
When both $p$ and $\rho$ are known, then $\mathcal{R}_n(C_n^*, p, \rho) \approx \mathcal{M}^*(n, p, \rho) - R(p,d,\rho)$, where $C_n^* = (\phi_n^*, f_n^*, g_n^*)$ uses a near-optimal $d$-quantizer $\phi_n^* \in \mathcal{Q}_{d, \rho}$ from the infimum in $(\ref{modirateredun})$ and the binary encoder and decoder $(f_n^{*}, g_n^{*})$ are chosen such that the expected rate is within $1/n$ of the entropy per symbol. Hence, in this non-universal case, the problem of analyzing $\mathcal{R}_n(C_n, p, \rho)$ is reduced to determining how fast the expected rate of an optimal code converges to the rate-distortion function. In the universal case when $p$ is unknown, the first two terms on the right-hand side of $(\ref{helloremiwolf})$ quantify the \emph{price of universality}. Our first two results in this paper will demonstrate achievable bounds for the price of \textit{universal distortion}, whose exact framework is described next.  

In the \emph{universal distortion} setting, the modified formulation of a $d$-semifaithful block code $\tilde{C}_n$ is 
\begin{align}
\begin{split}
    \phi_n & : A^n \times \mathcal{D} \to B_{\phi_n} \subset B^n\\
    f_n & : B_{\phi_n}  \to \mathcal{B}^*\\
    g_n & : \mathcal{B}^* \to B_{\phi_n},
\end{split} \label{dsemi_unknown}
\end{align}
where $\phi_n$ is now a $d$-quantizer w.r.t. the input distortion measure. Thus the distortion measure is not known in advance and only revealed to the $d$-quantizer at run-time.
\begin{remark}
To contrast $(\ref{dsemi})$ and $(\ref{dsemi_unknown})$, let us temporarily assume that $\mathcal{D} = \{ \rho_1, \rho_2, ..., \rho_m \}$ consists of a finite number of distortion measures. Then, $(\ref{dsemi})$ is a special case of $(\ref{dsemi_unknown})$ with $m=1$. Moreover, a $d$-semifaithful code in $(\ref{dsemi})$ achieving a rate redundancy of $\mathcal{R}_n(C_n, p, \rho')$ for an arbitrary distortion measure $\rho'$ can be extended to a universal distortion code in $(\ref{dsemi_unknown})$ to achieve a rate redundancy of $\mathcal{R}_n(C_n, p, \rho) + \ln m / n$ for all $\rho \in \mathcal{D}$. This can be done by taking a union of the codebooks of the $m$ codes (call them $C_n^{(1)}, C_n^{(2)}, \ldots, C_n^{(m)}$, where $C_n^{(i)}$ is a standard $d$-semifaithful code for the distortion measure $\rho_i$). Then when $(x^n,\rho)$ is an input to the quantizer for some $\rho \in \mathcal{D}$, a two-stage binary encoder can encode $\phi_n(x^n, \rho)$ by first communicating the label $j \in \{1,2,\ldots, m \}$ of the codebook followed by using the binary encoder of $C_n^{(j)}$. In the general setting, $\mathcal{D}$ is infinite so this approach fails. 
\label{naiveremark}
\end{remark}

The main technical contributions of the paper are to show how to obtain universality over $\rho$ given that $\mathcal{D}$ is a continuous space, and then extend this universality over distortion constraint $d$ as well. The latter provides a generalization of the universal distortion framework in which both the distortion measure $\rho$ and the distortion constraint $d$ can be run-time inputs to the quantizer only. We will call this the \emph{generalized} universal distortion code $\tilde{C}_n$ which has the following formulation:  
\begin{align}
\begin{split}
    \phi_n & : A^n \times \mathcal{D} \times \mathbb{R}_{> 0} \to B_{\phi_n} \subset B^n\\
    f_n & : B_{\phi_n}  \to \mathcal{B}^*\\
    g_n & : \mathcal{B}^* \to B_{\phi_n},
\end{split} \label{dsemi_unknown2}
\end{align}

We now define the counterparts to $(\ref{rate_redunda})$ and $(\ref{modirateredun})$ for the two new frameworks in $(\ref{dsemi_unknown})$ and $(\ref{dsemi_unknown2})$. For a universal distortion code $\tilde{C}_n$ in $(\ref{dsemi_unknown})$, we simply redefine $(\ref{rate_redunda})$ and $(\ref{modirateredun})$ to include the distortion measure as an input to the $d$-quantizer; the rate redundancy is given by
\begin{align}
  &\mathcal{R}_n(\tilde{C}_n,p,\rho) \triangleq \frac{1}{n} \mathbb{E} \left [l\left( f_n\left(\phi_n(X^n, \rho)\right) \right) \ln 2 \right ] - R(p,d,\rho) \label{isola14}
  \end{align}
and the operational rate redundancy (or price of universal distortion) is
\begin{align}
  &\overline{\mathcal{R}}_n(\tilde{C}_n,p,\rho) \notag\\
  &\triangleq \frac{1}{n} \mathbb{E} \left [l\left( f_n\left(\phi_n(X^n, \rho)\right) \right) \ln 2 \right ] -  \inf_{\phi_n \in \mathcal{Q}_{d, \rho}} \frac{H(\nu_{p^n, \phi_n})}{n}. \label{cokapepsiadeel}
\end{align}
For the generalized universal distortion $d$-semifaithful code $\tilde{C}_n$, we define the rate redundancies to include the distortion constraint $d$ as an additional parameter: 
\begin{align*}
  \mathcal{R}_n(\tilde{C}_n,p,\rho, d) \triangleq \frac{1}{n} \mathbb{E} \left [l\left( f_n\left(\phi_n(X^n, \rho, d)\right) \right) \ln 2 \right ] - R(p,d,\rho)
\end{align*}
and 
\begin{align}
    &\overline{\mathcal{R}}_n(\tilde{C}_n,p,\rho,d) \notag\\
    &\triangleq \frac{1}{n} \mathbb{E} \left [l\left( f_n\left(\phi_n(X^n, \rho, d)\right) \right) \ln 2 \right ] -  \inf_{\phi_n \in \mathcal{Q}_{d, \rho}} \frac{H(\nu_{p^n, \phi_n})}{n}. \label{coola}
\end{align}

\section{Main Results}

Our first result establishes an achievable minimax convergence rate for the operational rate redundancy $\overline{R}_n(\tilde{C}_n, p, \rho,d)$ as defined in $(\ref{coola})$. The achievability scheme uses an approach which is based on VC dimension~\cite{yaser1}. It extends \cite[Lemma 4]{silva1} to the generalized universal distortion setting of $(\ref{dsemi_unknown2})$.

\begin{theorem}
In the generalized universal distortion setting,
\begin{align*}
    \limsup_{n \to \infty}\,\, \inf_{\tilde{C}_n } \,\, \sup_{\stackrel{ (p, \rho ) \in \mathcal{P}(A) \times \mathcal{D} }{d \in (0, \infty)} } \, \frac{\overline{R}_n(\tilde{C}_n, p, \rho, d)}{\ln n / n} \leq J^2 K^2 + J - 2,
\end{align*}
where the infimum is over all codes which meet the input distortion constraint with respect to the input distortion measure. 
\label{silvaext}
\end{theorem}
\noindent
\textit{Proof:} The proof is given in Section \ref{thmsilvaextproof}.
\\

The idea behind the proof is the following. The domain of a general quantizer $\phi_n$ is $A^n \times \mathcal{D} \times \mathbb{R}_{>0}$. We take inspiration from the fact that $A^n$ can be partitioned into a polynomial number of equivalence classes, namely type classes. Similarly, we can partition $\mathcal{D} \times \mathbb{R}_{>0}$ into a polynomial number of equivalence classes as follows. For each distortion measure $\rho \in \mathcal{D}$ and $d > 0$, define $h_{\rho, d}: \mathcal{P}_n(A \times B) \to \{-1, +1 \}$ to be a linear classifier dividing the space $\mathcal{P}_n(A \times B)$ into half-spaces as follows:
\begin{align}
    h_{\rho, d} (s) &= \begin{cases}
    +1 & \,\,\,\,\,\,\,\,\,\, \text{if} \,\,\,\,\,\sum_{j ,k} s(j, k) \rho(j, k) \leq d\\
    -1 & \,\,\,\,\,\,\,\,\,\, \text{if} \,\,\,\,\,\sum_{j ,k} s(j, k) \rho(j, k) > d.
    \end{cases} \label{equivfunc}
\end{align}
Let $\mathcal{H} = \{h_{\rho,d} : \rho \in \mathcal{D}, d > 0 \}$. We say that the two ordered pairs $(\rho^{(1)}, d^{(1)})$ and $(\rho^{(2)}, d^{(2)})$ are equivalent if $h_{\rho^{(1)}, d^{(1)}} = h_{\rho^{(2)}, d^{(2)}}$, i.e., 
\begin{align*}
    h_{\rho^{(1)}, d^{(1)}}(s) = h_{\rho^{(2)}, d^{(2)}}(s)
\end{align*}
for all $s \in \mathcal{P}_n(A \times B)$. This defines an equivalence relation on $\mathcal{D} \times \mathbb{R}_{> 0}$ and, therefore, partitions $\mathcal{D} \times \mathbb{R}_{> 0}$ into equivalence classes $\left \{[\mathcal{D}]_{\rho, d} : \rho \in \mathcal{D}, d > 0 \right \}$, where the equivalence class $[\mathcal{D}]_{\rho,d}$ is defined as
\begin{align*}
    [\mathcal{D}]_{\rho,d} \triangleq \left \{ (\rho',d' ) \in \mathcal{D} \times \mathbb{R}_{>0}: h_{\rho', d'} = h_{\rho, d} \right \}.
\end{align*}
Any two pairs $(\rho^{(1)}, d^{(1)})$ and $(\rho^{(2)}, d^{(2)})$ in the same equivalence class are operationally interchangeable for encoding and decoding purposes, i.e., 
\begin{align}
    \rho^{(1)}(x^n, y^n) \leq d^{(1)} \iff \rho^{(2)}(x^n, y^n) \leq d^{(2)} \label{supppropo}
\end{align}
for all $x^n \in A^n$ and $y^n \in B^n$. Note that $|\mathcal{P}_n(A \times B)| \leq (n+1)^{JK - 1}$. Each equivalence class $[\mathcal{D}]_{\rho,d}$ can be uniquely associated with the corresponding $h_{\rho,d}$ which can be uniquely associated with an $M$-tuple of $\pm 1$'s, also called a dichotomy on $\mathcal{P}_n(A \times B)$, where $M \leq (n+1)^{JK - 1}$. Therefore, the number of equivalence classes, call it $m_{\mathcal{H}}(n)$, is equal to the number of distinct dichotomies on $\mathcal{P}_n(A \times B)$ which can be generated by $\mathcal{H}$. Clearly, $m_{\mathcal{H}}(n) \leq 2^{(n+1)^{JK - 1}}$. However, the number of dichotomies which $\mathcal{H}$ can generate on $\mathcal{P}_n(A \times B)$ is limited by the VC dimension \cite[Definition 2.5]{yaser1} of $\mathcal{H}$. Since $\mathcal{H}$ is a set of linear classifiers in $JK$-dimensional space, the VC dimension of $\mathcal{H}$ is at most $JK+1$ \cite[4.11]{vapnik1}. Therefore, since the number of joint $n$-types is at most $(n+1)^{JK-1}$, the maximum number of dichotomies\footnote{For an exact number of dichotomies on points satisfying certain conditions, see \cite[Theorem 1]{4038449}.} generated by $\mathcal{H}$ is (see \cite[2.9]{yaser1} and \cite[2.10]{yaser1}) 
\begin{align*}
    m_{\mathcal{H}}(n) &\leq \sum_{i=0}^{JK+1} \binom{|\mathcal{P}_n(A \times B)|}{i}\\
    &\leq \left((n+1)^{JK-1}\right )^{JK+1} + 1\\
    &= (n+1)^{J^2K^2 - 1} + 1.
\end{align*}
Let $(\rho_1,d_1 ), (\rho_2, d_2 ), ..., (\rho_{m_{\mathcal{H}}(n)},d_{m_{\mathcal{H}}(n)} )$ be the representative distortion measures from the $m_{\mathcal{H}}(n)$ equivalence classes of $\mathcal{D} \times \mathbb{R}_{>0}$. These are the polynomial number of distortion measures we desire. The above discussion can be encapsulated in the following proposition. 
\begin{proposition} 
There are $ m_{\mathcal{H}}(n) \leq (n+1)^{J^2 K^2 - 1} + 1$ equivalence classes of $\mathcal{D} \times \mathbb{R}_{> 0}$, denoted by $[\mathcal{D}]_{\rho_1, d_1}$, $[\mathcal{D}]_{\rho_2, d_2}$, \ldots, $[\mathcal{D}]_{\rho_{m_{\mathcal{H}}(n)}, d_{m_{\mathcal{H}}(n)}}$. A $d$-semifaithful code $C_n$ with respect to a distortion measure $\rho$ is also $d'$-semifaithful with respect to distortion measure $\rho'$ for all $(\rho', d') \in [\mathcal{D}]_{\rho, d}$ in the same equivalence class.  
\label{propo2}
\end{proposition}

Our next result (Theorem \ref{specthm}) uses a quantization approach to reduce the continuum of distortion measures into a polynomial number of distortion measures. This approach leads to a better redundancy bound than in Theorem \ref{silvaext}. However, the result holds only for uniformly bounded distortion measures in $\mathcal{D}^{\rho_{\max}}$. The coding scheme uses a custom quantization of $\mathcal{D}^{\rho_{\max}}$ as a function of $d$ and a post-correction scheme to prove a minimax achievability result for $\overline{R}_n(\tilde{C}_n, p, \rho)$ as defined in $(\ref{cokapepsiadeel})$. In the low distortion regime, the coding scheme in Theorem \ref{specthm} requires a finer quantization of the space of distortion measures. Specifically, the lower order terms in the given redundancy bound entail an increasing penalty with decreasing $d$. Consequently, the result only applies to the universal distortion framework in $(\ref{dsemi_unknown})$, i.e., the redundancy bound does not hold uniformly over all distortion levels $d \in \mathbb{R}_{>0}$.

Consider a quantization $\mathcal{D}_n^{q}$ of $\mathcal{D}^{\rho_{\max}}$, which is parametrized by some integer $q$: 
\begin{definition} A distortion measure $\rho \in \mathcal{D}_n^q \subset \mathcal{D}^{\rho_{\max}}$ if for all $j \in A$, $k \in B$, we have $\rho(j, k) = m\, \rho_{\max} / (q n)$ for some integer $m$ satisfying $0 \leq m \leq q n$. \label{defo1par}
\end{definition}
\begin{definition}
Given the distortion measure $\rho \in \mathcal{D}^{\rho_{\max}}$, we will denote by $[\rho] \in \mathcal{D}_n^q$ the quantization of $\rho$ which satisfies 
\begin{itemize}
    \item $[\rho](j, k) \leq \rho(j, k)$ for all $j \in A$, $k \in B$. 
    \item $\big | [\rho](j, k) - \rho(j, k) \big | < \frac{\rho_{\max}}{qn}$ for all $j \in A$, $k \in B$. 
\end{itemize} 
\label{defo2param}
\end{definition}
\noindent

\begin{theorem}
For any $d > 0$, there exists a universal distortion $d$-semifaithful code $\tilde{C}_n$ satisfying 
\begin{align*}
    \limsup_{n \to \infty}\,\, \sup_{(p, \rho ) \in \mathcal{P}(A) \times \mathcal{D}^{\rho_{\max}} } \, \frac{\overline{R}_n(\tilde{C}_n, p, \rho)}{\ln n / n} \leq J K + J.
\end{align*}
\label{specthm}
\end{theorem}
\noindent
\textit{Proof:} The proof is given in Section \ref{specthmproof}.

So far, we have given results establishing convergence to 
\begin{align}
    \inf_{\phi_n \in \mathcal{Q}_{d, \rho} } \frac{H(\nu_{p^n, \phi_n})}{n} \label{targetlow}
\end{align}
instead of the rate-distortion function. The operational nature of $(\ref{targetlow})$ made it an easier target: one did not need to establish a single-letter characterization of the performance of the proposed codes. Establishing convergence to the rate-distortion function in the universal distortion setting involves various technical challenges related to continuity, smoothness, $d$-ball covering and convergence of $\mathbb{E}_p\left[R(T,d,\rho) \right]$ to $R(p,d,\rho)$. Both the VC dimension approach (Theorem \ref{silvaext}) and the quantization approach (Theorem \ref{specthm}) establish that a polynomial number of distortion measures suffice for achieving universality over the continuous space $\mathcal{D}^{\rho_{\max}}$. Hence, as alluded to in Remark \ref{naiveremark}, one approach could be to take a standard universal $d$-semifaithful code from previous works which works for a fixed arbitrary distortion measure and instantiate it a polynomial number of times. Using this idea with the VC dimension approach and the quantization approach would add penalties of $(J^2 K^2 - 1) \ln n / n$ and $JK \ln n / n$ to the rate redundancy, respectively. If one seeks optimal pre-log factors, then such an approach is unlikely to succeed.        

In lossless coding, Rissanen \cite{rissanen1} established an optimal lossless coding rate redundancy of $0.5 k \ln n/ n$ for most sources in a parametric class, where $k$ is the dimension of the parametric space. In universal lossy coding, the corresponding parametric space may at first seem to be the set of all distributions on the source alphabet, which has dimension $J-1$. However, the rate-distortion function for an i.i.d.\ source $p$ has the following alternative characterization (see, e.g., \cite{650987lp}):
\begin{align}
    R(p,d,\rho) &= \inf_{q \in \mathcal{P}(B)} R(q, p,d,\rho) \label{altrd} 
\end{align}
where the infimum is over all probability distributions $q$ on the reproduction alphabet, and $R(q, p,d,\rho)$ is the rate achieved by a random codebook used to compress the source data within distortion $d$ w.r.t. distortion measure $\rho$, where the codewords are randomly generated i.i.d. according to $q$. Hence, for each distribution $p$ over the source alphabet, there is a corresponding optimal distribution $q=Q^{p,d,\rho}$ on the reconstruction alphabet which achieves the rate-distortion function. As $p$ varies over the simplex of probability distributions, $Q^{p,d,\rho}$ varies over a space of dimension $K-1$. It would seem to be the dimension of the space of distributions on the reconstruction alphabet which should determine the coefficient before $\ln n/ n$, since it captures all of the distributional information that will be revealed to the decoder. Indeed, for the fixed-rate variant of traditional universal lossy coding, \cite{yang2} established an optimal (assuming $K \leq J$) pointwise distortion redundancy of 
\begin{align}
       \left( \frac{K}{2}\right) \frac{\ln n}{n} \Big | \frac{\partial }{\partial R} d(p,R,\rho) \Big | + o \left(\frac{\ln n}{n} \right) \label{2nonnon}
\end{align}
for source distributions $p$ satisfying certain regularity conditions, where the distortion redundancy is defined as the difference between expected distortion of the code and the distortion-rate function. When contrasted with the optimal distortion redundancy, given by \cite[Theorem 1]{yang1}
\begin{align}
      \left(\frac{1}{2} \right)\frac{ \ln n}{n} \Big | \frac{\partial }{\partial R} d(p,R,\rho) \Big |  + o \left(\frac{\ln n}{n} \right),
     \label{2non}
\end{align}
of non-universal fixed rate coding, we see that $(K-1)/2$ is the "price of universality", which is consistent with $(\ref{altrd})$ and Rissannen's redundancy result. For traditional universal $d$-semifaithful codes, the optimal rate redundancy is not precisely characterized; for source distributions $p \in \mathcal{P}_d \subset \mathcal{P}(A)$ satisfying certain regularity conditions,  a converse result is known \cite[Theorem 1]{yang3} giving a lower bound of $K/2 \ln n / n + o(\ln n / n)$ for the rate redundancy for most $p \in \mathcal{P}_d$ while an achievability result \cite[Theorem 2]{yang3} of $(K+2)/2 \ln n / n$ is given for all $p \in \mathcal{P}_d$.

In the universal-distortion setup considered in this paper, the variation of distortion measure $\rho$ does not change the parametric space $\mathcal{P}(B)$ of the formulation in $(\ref{altrd})$. Hence, $[K/2, (K+2)/2]$ is a reasonable guess for the range of optimal pre-log factors for the rate redundancy $\mathcal{R}_n(\tilde{C}_n,p,\rho )$ as defined in $(\ref{isola14})$. Our final result gives a universal-distortion achievability result with the pre-log factor within this range. We consider a subset  $\mathcal{S}_{d} \subset \mathcal{P}(A) \times \mathcal{D}^{\rho_{\max}}$ defined as follows:
\begin{definition}
For all $(p, \rho) \in \mathcal{S}_{ d}$, we have   
\begin{enumerate}
    \item $p(j) >0 $ for all $j \in A$,
    \item $Q^{p,d,\rho}$ is unique,  
    \item $Q^{p,d,\rho}(k) > 0$ for all $k \in B$, and 
    \item $0 < d < \min_{k \in B} \sum_{j \in A} p(j) \rho(j, k)$. 
\end{enumerate}
\label{defineSdset}
\end{definition}

\begin{remark} The uniqueness\footnote{A non-trivial sufficient condition for the uniqueness of $Q^{p,d,\rho}$ for full-support $p$ can be found in \cite[Lemma 7]{yang2}.} of the optimal output distribution $Q^{p,d,\rho}$ implies the uniqueness of the optimal channel $W^*_{B|A}[p,d,\rho]$ by the following well-known relation (see, e.g., Equation $(10.124)$ in \cite{thomas_cov}):
\begin{align*}
    W^*_{B|A}[p,d,\rho](k|j) = \frac{Q^{p,d,\rho}(k) \exp \left(- \lambda^* \rho'(j,k)  \right)}{\sum \limits_{k' \in B} Q^{p,d,\rho}(k') \exp \left( -\lambda^* \rho'(j,k') \right) },
\end{align*}
which holds for all $(p, \rho) \in \mathcal{S}_d$ and
where $-\lambda^* = \partial R(p,d,\rho) / \partial d$. The uniqueness of $Q^{p,d,\rho}$ is a common assumption in past works \cite{yang1}, \cite{yang2}, \cite{yang3}, \cite{kontoyiannis2}. Also see \cite[Remark 1]{yang2}, \cite[Remark b), p. 2283]{kontoyiannis2}, \cite[Remark, p. 817]{yang1} for discussion and examples of $(p,\rho)$ satisfying the full-support assumptions on $p$ and $Q^{p,d, \rho}$. The fourth condition in Definition \ref{defineSdset} restricts our attention to the interesting case where $R(p,d,\rho) > 0$.    
\end{remark}

\begin{remark}
Under the conditions of Definition \ref{defineSdset}, we have a $o(\ln n / n)$ convergence of $\mathbb{E}_p \left [ R(T,d,\rho) \right]$ to $R(p,d,\rho)$ from above. This result is implicit in \cite{yang1} but we have included a proof in Appendix \ref{lemmataylorproof} for convenience. 
\label{hepzivb}
\end{remark}
\begin{theorem}
Fix any $d > 0$.  There exists a random $d$-semifaithful code $\tilde{C}_n = (\Phi_n, f_n, g_n)$ in the universal distortion setting such that for every $(p, \rho) \in \mathcal{S}_d$, 
\begin{align*}
    \mathbb{E}_c \left [ \mathcal{R}_n(\tilde{C}_n, p, \rho) \right] \leq \left( \frac{K}{2}  + 1\right) \frac{\ln n}{n} +  o\left( \frac{\ln n}{n}\right) ,
\end{align*}
where $\mathbb{E}_c\left [\cdot \right]$ denotes expectation with respect to the random code. 
\label{thmyuext}
\end{theorem}
\textit{Proof:} The proof is given in Section \ref{thmyuextproof}.

\begin{remark}
Unlike Theorems \ref{silvaext} and \ref{specthm}, the convergence in Theorem \ref{thmyuext} is pointwise in $p$ and $\rho$.
\end{remark}

As discussed above, we use a random coding argument to prove Theorem \ref{thmyuext}. Since $p$ is unknown and $\rho$ is not known at design time either, the encoder and decoder share a random codebook with i.i.d. codewords from the normalized maximum-likelihood (NML) distribution over the reconstruction alphabet (see the definition of NML distribution in $(\ref{tutu})$). Then given the type $t$ and distortion measure $\rho$ at runtime, the encoder uses acceptance-rejection sampling from the codebook to obtain i.i.d. codewords according to $(Q^{t,d,\rho})^n$ and sends the index of the first one meeting the distortion constraint. This is different from the random coding argument in \cite{yang2, yang3} which uses "$1/\sqrt{n}$-type" quantization of the space $\mathcal{P}(B)$ and generates uniform samples from each type. Arguably, our approach circumvents the need for more delicate continuity and smoothness arguments with respect to $Q^{t,d,\rho}$. See also \cite{https://doi.org/10.48550/arxiv.2203.03305} for another variant of the random coding argument in which the shared random codebook has codewords drawn from a uniform mixture of i.i.d. sources on the reconstruction alphabet.   

At the heart of our random coding argument is Lemma \ref{dballlemma} which bounds the probability of a codeword $Y^n \sim (Q^{t,d,\rho})^n$ meeting the distortion constraint with a type $t$ source sequence. 
 
Define for any $\delta > 0$, 
\begin{align}
    \mathcal{N}_\delta(p, \rho) \triangleq \{(t,\rho') \in \mathcal{P}_n(A) \times \mathcal{D}^{\rho_{\max}} : ||t-p||_2 < \delta \text{ and } ||\rho - \rho'||_F < \delta \}. \label{o(1)neigh}
\end{align}

\begin{lemma}
Fix any $d > 0$ and $(p, \rho) \in \mathcal{S}_d$. Then there exists a $\delta > 0$ such that   
\begin{align}
    &\mathbb{P}(\rho'_n(x^n, Y^n) \leq d) \notag \\
    &\geq  \exp \left( -n R(t,d,\rho')  - \frac{1}{2} \ln n + O(1) \right) \label{crucialO(1)}
\end{align}
for all $(t,\rho') \in \mathcal{N}_\delta(p, \rho)$,
where $x^n \in T^n_A(t)$, $Y^n$ is i.i.d. according to $Q^{t,d,\rho'}$ and the $O(1)$ term depends only\footnote{We do not mention the dependence on alphabet sizes since those are fixed throughout the paper.} on $p,d, \rho$ and $\rho_{\max}$.     
\label{dballlemma}
\end{lemma}
\noindent
\textit{Proof of Lemma \ref{dballlemma}:} The proof of lemma \ref{dballlemma} is given in Appendix \ref{dballlemmaproof}.

Different variants of the lower bound in Lemma \ref{dballlemma} underlie the random coding approaches used in previous works to prove achievability schemes; see, e.g., \cite[Lemma 3]{yang1}, \cite[Lemma 1]{https://doi.org/10.48550/arxiv.2203.03305}. The main idea is to have a random codebook of i.i.d. codewords $\{Y_1^n, Y_2^n, Y_3^n, \ldots  \}$ available to both the encoder and decoder. Then given an input source sequence $x^n$ to compress, the encoder conveys to the decoder the index $I$ of the first codeword $Y_I^n$ which meets the distortion constraint with $x^n$. The smaller the value of $I$, the shorter the length of the binary encoding. It is easy to see that $I$ is a geometric random variable with success probability lower bounded as in $(\ref{crucialO(1)})$. A useful property of Lemma \ref{dballlemma} is that the "$O(1)$" term in $(\ref{crucialO(1)})$ is uniformly bounded over a neighborhood $\mathcal{N}_\delta(p, \rho)$; in particular, it is independent of the type $t = t(x^n)$ of the source sequence $x^n$ which facilitates the expected rate analysis in the proof of Theorem \ref{thmyuext}. A second advantage is that since the lower bound in $(\ref{crucialO(1)})$ holds uniformly over all types and distortion measures in a neighborhood, one can show that with high probability, the worst-case (i.e., maximum) integer index $I$ over all types and distortion measures is small. This argument is made rigorous in the proof of Proposition \ref{instanceofthm3from} and is a crucial part in our next discussion about obtaining a deterministic code.             

Our Theorem \ref{thmyuext} can be viewed as a partial strengthening of \cite[Theorem 2]{yang3} in that our result is in the more general universal distortion setting and has fewer regularity conditions which are actually a subset of the regularity conditions used in \cite[Theorem 2]{yang3}. However, Theorem \ref{thmyuext} only proves the existence of a random code while \cite[Theorem 2]{yang3} gives a deterministic code. Below, we outline a method to \textit{derandomize} our code in Theorem \ref{thmyuext} by fusing ideas from \cite{yang3} and \cite{https://doi.org/10.48550/arxiv.2202.04481}, but this comes at the expense of making our regularity conditions in Definition \ref{defineSdset} stricter to match those in \cite{yang3}. 

The proofs of both theorems (Theorem \ref{thmyuext} and \cite[Theorem 2]{yang3}) begin with a random coding argument. As such, both proofs rely on lower bounding the probability of a random codeword meeting the distortion constraint. The result in \cite{yang3} obtains this bound by estimating the size of the $d$-ball around any given type $t$ source sequence (see \cite[Lemma 3]{yang1}) using a technical counting argument given in \cite[Appendix]{yang1}. On the other hand, we use standard large deviations techniques and the Berry-Esseen Theorem (Lemma \ref{dballlemma} in this paper). An advantage of our method is that it is more easily extended to variable distortion measures; as remarked earlier, our lower bound holds uniformly in a neighborhood around any given $(p, \rho)$. In view of this, it is possible to show strong universality or uniform convergence over the neighborhood around any given $(p, \rho)$, similar to how \cite[Theorem 2]{yang2} or \cite[Theorem 2]{yang3} showed strong universality over the neighborhood around $p$. Hence, we have the following proposition:
\begin{proposition}
Fix any $d > 0$ and $(p, \rho) \in \mathcal{S}_d$. There exist a neighborhood $\mathcal{N}_\delta(p, \rho)$ for some $\delta > 0$ and a (deterministic) universal distortion $d$-semifaithful code $\tilde{C}_n = (\phi_n, f_n, g_n)$ satisfying 
\begin{align}
    \sup_{(p',\rho') \in \mathcal{N}_\delta(p, \rho)} \left [ \frac{\ln 2}{n} \mathbb{E}_{p'}\left[l(f_n(\phi_n(X^n, \rho'))) \right] - \mathbb{E}_{p'}\left [ R(T,d,\rho') \right] \right ] \leq \left(\frac{K}{2} + 1 \right)\frac{\ln n}{n} + O\left( \frac{\ln \ln n}{n} \right). \label{boskfoens}
\end{align}
\label{instanceofthm3from}
\end{proposition}
\textit{Proof:} The proof is given in Appendix \ref{instanceofthm3fromproof}.

Unlike Theorem \ref{thmyuext}, whose achievability bound holds pointwise for each $p$ and $\rho$, the bound in $(\ref{boskfoens})$ holds uniformly for $(p', \rho')$ in a neighborhood around a given $(p, \rho)$. In other words, we went from weak universality to strong universality at the expense of shrinking the set of $(p, \rho)$ over which universality is achieved. However, this strong universality or uniform convergence allows us to obtain a deterministic code in Proposition \ref{instanceofthm3from} as opposed to a random code in Theorem \ref{thmyuext}; in the proof of Proposition \ref{instanceofthm3from}, we used a union bound over types and equivalence classes of distortion measures over a suitable neighborhood around $(p, \rho)$. The notion of equivalence classes of distortion measures is a straightforward corollary of Proposition \ref{propo2}: for a given blocklength $n$ and distortion level $d$, there are $ \xi \leq (n+1)^{J^2 K^2 - 1} + 1$ equivalence classes of $\mathcal{D}^{\rho_{\max}}$, denoted by $[\mathcal{D}]_{\rho_1}$, $[\mathcal{D}]_{\rho_2}$, \ldots, $[\mathcal{D}]_{\rho_{\xi}}$, where $\rho_1, \rho_2, \ldots, \rho_\xi$ are arbitrarily chosen representative distortion measures. A $d$-semifaithful code with respect to a distortion measure $\rho$ is also $d$-semifaithful with respect to all distortion measures $\rho' \in [\mathcal{D}]_{\rho}$ in the same equivalence class.

Now, as mentioned before, under certain additional assumptions\footnote{Similar assumptions can be found in \cite[p. 129]{yang2}} on $p$ and $\rho$ from \cite[p. 8]{yang3}, it can be shown that the set $\mathcal{S}_d$ in Definition \ref{defineSdset} is an open set. Hence, $\mathcal{S}_d$ can be expressed as a countable union of compact subsets, each of which (by Heine-Borel theorem) can be covered by a finite union of neighborhoods of the form $\mathcal{N}_{\delta}(p, \rho)$. Then with a similar argument as in \cite[Theorem 2 (ii)]{yang3} or \cite[Corollary 2]{yang2}, the existence of a deterministic, universal-distortion $d$-semifaithful code whose expected rate converges pointwise to $\mathbb{E}_p[R(T,d,\rho)]$ for all $(p, \rho) \in \mathcal{S}_d$ can be proved. Subsequently, an application of Lemma \ref{lemmataylor} in Appendix \ref{lemmataylorproof} guarantees a $o(\ln n / n)$ convergence of $\mathbb{E}_p[R(T,d,\rho)] \to R(p,d,\rho)$. We omit this result and instead keep Theorem \ref{thmyuext} as one of our main results to keep the definition of $\mathcal{S}_d$ simpler and the associated regularity conditions slightly more general.

\section{Concluding Remarks}

The three main results show how to achieve universal distortion using three different approaches: the VC dimension approach, the quantization approach and the random coding approach. None of the results, however, show minimax convergence to the rate-distortion function. From
Theorem~\ref{silvaext}, we have that the redundancy $\overline{R}_n(\tilde{C}_n, p, \rho,d )$
can be made to vanish uniformly in $p,d$ and $\rho$ for some code. Thus to 
obtain convergence to the rate-distortion function that is uniform over $p$,
it suffices to show that
\begin{equation}
    \label{eq:uniform1}
    \lim_{n \rightarrow \infty} \sup_{p}  \Big |\inf_{\phi_n \in \mathcal{Q}_{d, \rho} } \frac{H(\nu_{p^n, \phi_n})}{n}
      - R(p,d,\rho) \Big | = 0 \quad \text{for all $\rho$ and $d > 0$}.
\end{equation}
A pointwise version of this result,
\begin{equation}
    \lim_{n \to \infty} \inf_{\phi_n \in \mathcal{Q}_{d, \rho} } \frac{H(\nu_{p^n, \phi_n})}{n} = R(p,d,\rho) 
      \quad \text{for all $p,\rho$ and $d$}, \label{kuta}
\end{equation}
is known (cf.~\cite[Theorem 4]{kieffer1}). One approach is thus to strengthen 
(\ref{kuta}) to (\ref{eq:uniform1}), perhaps to include uniformity over $\rho$ and $d$ as well. 

An alternative is to show
that the gap between expected codeword length and the code-independent 
quantity  $\mathbb{E}_p [ R(T,d,\rho)]$ (where $T$ is the $n$-type of a
source string generated i.i.d.\ according to $p$) vanishes uniformly
and then extend Lemma~\ref{lemmataylor} to show
\begin{equation}
    \lim_{n \rightarrow \infty} \sup_{p} \left(
      \mathbb{E}_p [ R(T,d,\rho)] - R(p,d,\rho\right))^+ = 0,
      \label{eq:uniform3}
\end{equation}
where $x^+ = \max(x,0)$. The following lemma, which is proven
in Appendix \ref{lemmagoldenproof}, shows that (\ref{eq:uniform3})
is in fact necessary.
 
\begin{lemma}
For all $n \in \mathbb{N}$, any $d$-semifaithful code $C_n = (\phi_n, f_n, g_n)$ satisfies
\begin{align*}
    &\frac{1}{n} \mathbb{E} \left [l\left( f_n\left(\phi_n(X^n, \rho)\right) \right) \ln 2 \right ] \geq\\
    &\,\,\,\,\,\,\,\, \mathbb{E}_p \left [ R(T,d,\rho) \right] - (J K  + J - 2) \frac{\ln n}{n} - \frac{JK + J - 2}{n}
\end{align*}
for all $p \in \mathcal{P}(A)$ and $\rho \in \mathcal{D}$. 
\label{lemmagolden}
\end{lemma}  

Lemma \ref{lemmagolden} shows that the emergence of the quantity 
$\mathbb{E}_p [ R(T,d,\rho)]$ is not an artifact of the proof of 
Theorem~\ref{thmyuext} or most other theorems \cite{yu1}, \cite{https://doi.org/10.48550/arxiv.2203.03305}, \cite{yang1} showing achievability results with respect to the rate-distortion function. See \cite[Theorem 2]{https://doi.org/10.48550/arxiv.2203.03305} for another converse result which gives a pointwise lower bound to the encoding length in terms of $R(t,d,\rho)$ where $t = t(x^n)$ is the type of any given realization of the source sequence. Indeed, \cite{https://doi.org/10.48550/arxiv.2202.04481} gives a minimax convergence to the rate-distortion function by first showing that the difference between the expected rate of an optimal code and $\mathbb{E}_p\left [ R(T,d,\rho)\right ]$ goes to zero uniformly over $p$ and $\rho$.

\section{Proof of Theorem \ref{silvaext} \label{thmsilvaextproof}}

From Proposition \ref{propo2}, there are a polynomial number of equivalence classes of $\mathcal{D} \times \mathbb{R}_{> 0}$. Let us focus first on one equivalence class $[\mathcal{D}]_{\rho_i, d_i}$ and a type $t \in \mathcal{P}_n(A)$. For each type $t$, let $p_t = \text{Unif} \left( T^n_A(t) \right)$ be the uniform distribution over the type class. Fix any $\epsilon > 0$. Given any equivalence class $[\mathcal{D}]_{\rho_i, d_i}$ and type $t$, it is always possible to choose $(\rho_i^t,d_i^t ) \in [\mathcal{D}]_{\rho_i, d_i}$ and $d_i^t$-quantizer $\phi_n^{t, i}$ with respect to $\rho_i^t$ such that  
\begin{align*}
    \frac{H(\nu_{p_t, \phi_n^{t,i}})}{n} \leq  \inf_{\phi_n \in \mathcal{Q}_{d_i,\rho_i}} \frac{H(\nu_{p_t, \phi_n}) + \epsilon}{n}. 
\end{align*}
We now construct a $d_i^t$-semifaithful code $C_n^{t, i} = ( \phi_n^{t, i},f_n^{t, i}, g_n^{t, i}  )$ with respect to $\rho_i^t$ whose expected rate with respect to $p_t$ is given by
\begin{align}
    &\frac{1}{n}\mathbb{E}_{p_t} \left [ l(f_n^{t, i}(\phi_n^{t, i}(X^n))) \ln (2) \right ] \notag\\
    &\leq \frac{H(\nu_{p_t, \phi_n^{t,i}})}{n} + \frac{\ln(2)}{n} \notag \\
    &\leq  \inf_{\phi_n \in \mathcal{Q}_{d_i,\rho_i}} \frac{H(\nu_{p_t, \phi_n}) + \ln(2) + \epsilon}{n}, \label{sec1}
\end{align}
where the binary encoder and decoder $(f_n^{t, i}, g_n^{t, i})$ are chosen optimally such that the average expected length of the binary string is within $1/n$ of the entropy per symbol; see \cite[Theorem 5.4.1 and 5.4.2]{thomas_cov}. Hence, we have a $d_i^t$-semifaithful code $C_n^{t, i}$ for each type $t$ and distortion measure $\rho_i^t$, where $1 \leq i \leq m_{\mathcal{H}}(n)$.    

We now construct a generalized universal distortion code $\tilde{C}_n =  (\phi_n, f_n, g_n)$ by collecting all the previous codes. For any input source sequence $x^n$,  input distortion measure $\rho \in \mathcal{D}$ and input distortion constraint $d > 0$, let $t = t(x^n)$ be the type of $x^n$ and let $i$ be the integer such that $(\rho, d) \in [\mathcal{D}]_{\rho_i^t, d_i^t}$, where $1 \leq i \leq m_{\mathcal{H}}(n)$. The mapping of the $d$-quantizer $\phi_n$ is given by 
\begin{align*}
    \phi_n(x^n, \rho) = \phi_n^{t, i}(x^n),
\end{align*}
which satisfies the distortion constraint according to Proposition \ref{propo2} and $(\ref{supppropo})$. The encoder $f_n$  first sends 
\begin{align}
    &\log \left( |\mathcal{P}_n(A)| m_{\mathcal{H}}(n)   \right) + 1 \notag\\
    &\leq \log \left( (n+1)^{J-1} \left( (n+1)^{J^2 K^2 - 1} + 1 \right) \right) + 1 \notag \\
    &= (J^2 K^2 + J - 2) \log(n) + J^2 K^2 + J \label{first1}
\end{align}
bits to identify the type $t$ and equivalence class $i$ followed by the binary encoding $f_n^{t, i}(\phi_n^{t, i}(x^n))$. Therefore, the expected rate of this scheme is given by
\begin{align}
    &\frac{1}{n} \mathbb{E}_p \left [ l(f_n(\phi_n(X^n, \rho))) \ln(2) \right ]\notag \\ 
    &\leq (J^2 K^2 + J - 2) \frac{\ln(n)}{n} + \frac{\ln(2) (J^2 K^2 + J)}{n} \notag \\
    & \,\,\,\,\,\,\,\,\,\,\,\,\,\,+ \frac{1}{n}\mathbb{E}_p \left [ l(f_n^{T, i}(\phi_n^{T, i}(X^n))) \ln(2) \right ] \label{kuk}\\
    &= (J^2 K^2 + J - 2) \frac{\ln(n)}{n} + \frac{\ln(2) (J^2 K^2 + J)}{n} + \notag \\
    & \,\,\,\frac{1}{n} \sum_{t \in \mathcal{P}_n(A)} p^n(T^n_A(t)) \mathbb{E}_{p} \left [  l(f_n^{t, i}(\phi_n^{t, i}(X^n))) \ln(2) | t(X^n) = t \right ] \label{kukas} \\
    &\leq (J^2 K^2 + J - 2) \frac{\ln(n)}{n} + \frac{\ln(2) (J^2 K^2 + J)}{n} + \notag \\
    & \,\,\,\sum_{t \in \mathcal{P}_n(A)} p^n(T^n_A(t)) \left [\,\, \inf_{\phi_n \in \mathcal{Q}_{d_i,\rho_i}} \frac{H(\nu_{p_t, \phi_n}) + \ln(2) + \epsilon}{n}\,\, \right ] \label{hop}\\
    &= (J^2 K^2 + J - 2) \frac{\ln(n)}{n} + \frac{\ln(2) (J^2 K^2 + J)}{n} +  \notag \\
    & \,\,\, \sum_{t \in \mathcal{P}_n(A)} p^n(T^n_A(t)) \left [\,\, \inf_{\phi_n \in \mathcal{Q}_{d,\rho}} \frac{H(\nu_{p_t, \phi_n}) + \ln(2) + \epsilon}{n}\,\, \right ] \label{hop2}  \\
    &= (J^2 K^2 + J - 2) \frac{\ln(n)}{n} + \frac{\ln(2) (J^2 K^2 + J)}{n} + \frac{\ln (2) + \epsilon}{n} \notag \\
    & \,\,\,\,\,\,\,\,\,\,\,\,\,\,\,\,\,\,\,\,\,\,\,\,\,\,\,\,\,\,+ \mathbb{E}_p \left [\,\, \inf_{\phi_n \in \mathcal{Q}_{d,\rho}} \frac{H(\nu_{p_T, \phi_n}) }{n}\,\, \right ] \notag \\
    &\leq (J^2 K^2 + J - 2) \frac{\ln(n)}{n} + \frac{\ln(2) (J^2 K^2 + J)}{n} + \frac{\ln (2) + \epsilon}{n} \notag\\
    & \,\,\,\,\,\,\,\,\,\,\,\,\,\,\,\,\,\,\,\,\,\,\,\,\,\,\,\,\,\,\,\,\,\,\,\,\,\,\,\,\,\,\,\,\,\,\,\,\,\,\,\,\,\,\,\,\,\,\,\,\,\,\,\,\,\,\,\,\,\,\,\,\,\,+ \inf_{\phi_n \in \mathcal{Q}_{d,\rho}} \frac{H\left ( \mathbb{E}_p \left [ \nu_{p_T, \phi_n} \right ] \right ) }{n} \label{concav}\\
    &= (J^2 K^2 + J - 2) \frac{\ln(n)}{n} + \frac{\ln(2) (J^2 K^2 + J)}{n} + \frac{\ln (2) + \epsilon}{n} \notag \\
    & \,\,\,\,\,\,\,\,\,\,\,\,\,\,\,\,\,\,\,\,\,\,\,\,\,\,\,\,\,\,\,\,\,\,\,\,\,\,\,\,\,\,\,\,\,\,\,\,\,\,\,\,\,\,\,\,\,\,\,\,\,\,\,\,\,\,\,\,\,\,\,\,\,\,+ \inf_{\phi_n \in \mathcal{Q}_{d,\rho}} \frac{H\left (  \nu_{p^n, \phi_n}  \right ) }{n}. \label{vtdef}
\end{align}
In the last term of $(\ref{kuk})$, $T = t(X^n)$ is a random type. In $(\ref{hop})$, we use $(\ref{sec1})$ along with the fact that $X^n$ is i.i.d. according to $p$ and that conditioned on the type, $X^n$ is uniformly distributed over the type class. In $(\ref{hop2})$, we use the fact that $(\rho, d)$ and $(\rho_i,d_i)$ belong to the same equivalence class. In $(\ref{concav})$, we use concavity and Jensen's inequality. Finally, in $(\ref{vtdef})$, we use the definition of $\nu_{p_t, \phi_n}$ from $(\ref{nudef})$:
\begin{align*}
     \nu_{p_t, \phi_n}(\tilde{y}^n) &= \sum_{x^n \in A^n} p_t(x^n) \mathds{1} \left( \phi_n(x^n) = \tilde{y}^n \right)\\
     &= \sum_{x^n \in T^n_A(t)} p_t(x^n) \mathds{1} \left( \phi_n(x^n) = \tilde{y}^n \right)
\end{align*}
and 
\begin{align*}
    &\mathbb{E}_p \left [\nu_{p_T, \phi_n}(\tilde{y}^n) \right ] \\
    &= \sum_{t \in \mathcal{P}_n(A)} p^n(T^n_A(t))  \sum_{x^n \in T^n_A(t)} p_t(x^n) \mathds{1} \left( \phi_n(x^n) = \tilde{y}^n \right)\\
    &= \sum_{t \in \mathcal{P}_n(A)}  \sum_{x^n \in T^n_A(t)} \underbrace{p^n(T^n_A(t))  p_t(x^n)}_{ = p^n(x^n)} \mathds{1} \left( \phi_n(x^n) = \tilde{y}^n \right)\\
    &= \sum_{x^n \in A^n} p^n(x^n) \mathds{1} \left( \phi_n(x^n) = \tilde{y}^n \right)\\
    &= \nu_{p^n, \phi_n}(\tilde{y}^n).
\end{align*}
The upper bound in $(\ref{vtdef})$ holds uniformly over all $p$, $\rho$ and $d > 0$ which enables us to write $(\ref{vtdef})$ as
\begin{align*}
    \sup_{p, \rho, d}\, \overline{R}_n(\tilde{C}_n, p, \rho) &\leq (J^2 K^2 + J - 2) \frac{\ln n}{n} + O(n^{-1})
\end{align*}
Dividing both sides by $\ln n / n$ and taking the limit establishes the result of Theorem \ref{silvaext}.

\section{Proof of Theorem \ref{specthm}  \label{specthmproof}}

Let $q = \lceil \rho_{\max} / d \rceil $ and quantize $\mathcal{D}^{\rho_{\max}}$ with $\mathcal{D}_n^q$. For each type $t$, let $p_t = \text{Unif}(T^n_A(t))$ be the uniform distribution over the type class. Fix any $\epsilon > 0$. For each type $t \in \mathcal{P}_n(A)$ and $[\rho] \in \mathcal{D}_n^q$, it is always possible to choose a $d$-quantizer $\phi_n^{t, [\rho]}$ with respect to $[\rho]$ such that 
\begin{align*}
    \frac{H\left (\nu_{p_t, \phi_n^{t, [\rho]}}\right)}{n} \leq \inf_{\phi_n \in \mathcal{Q}_{d, [\rho]}} \frac{ H(\nu_{p_t, \phi_n}) + \epsilon}{n}.
\end{align*}
Hence, for each type $t$ and $[\rho] \in \mathcal{D}_n^q$, we can construct a $d$-semifaithful code $C_n^{t, [\rho]} = \left (\phi_n^{t, [\rho]}, f_n^{t, [\rho]}, g_n^{t, [\rho]} \right )$ with respect to $[\rho]$ whose expected rate with respect to $p_t$ is given by 
\begin{align}
    &\frac{1}{n}\mathbb{E}_{p_t} \left [ l(f_n^{t, [\rho]}(\phi_n^{t, [\rho]}(X^n))) \ln (2) \right ] \notag \\
    &\,\,\,\,\,\,\,\,\,\,\,\,\,\,\,\,\leq \frac{H(\nu_{p_t, \phi_n^{t,[\rho]}})}{n} + \frac{\ln(2)}{n} \notag \\
    &\,\,\,\,\,\,\,\,\,\,\,\,\,\,\,\,\leq \inf_{\phi_n \in \mathcal{Q}_{d, [\rho]}} \frac{ H(\nu_{p_t, \phi_n}) +\ln(2) +  \epsilon}{n}, \label{kutee11}
\end{align}
where the binary encoder and decoder $(f_n^{t, [\rho]}, g_n^{t, [\rho]})$ are chosen optimally such that the average expected length of the binary string is within $1/n$ of the entropy per symbol; see \cite[Theorems 5.4.1 and 5.4.2]{thomas_cov}.

We now construct a universal distortion $d$-semifaithful code $\tilde{C}_n =  (\phi_n, f_n, g_n)$ by using the previous codes 
$$\left \{ C_n^{t, [\rho]} : t \in \mathcal{P}_n(A), [\rho] \in \mathcal{D}_n^q \right \}$$ in conjunction with a post-correction scheme, which is described next. For any input source sequence $x^n$ and input distortion measure $\rho \in \mathcal{D}^{\rho_{\max}}$, let $t = t(x^n)$ be the type of $x^n$ and let $[\rho] \in \mathcal{D}_n^q$ be the appropriate quantization of $\rho$ as described in Definitions $\ref{defo1par}$ and $\ref{defo2param}$. The $d$-quantizer $\phi_n$ with respect to $\rho$ first uses $\phi_n^{t, [\rho]}$ to encode $x^n$ which satisfies $[\rho]_n(x^n, \phi_n^{t, [\rho]} (x^n)  ) \leq d$ which implies $\rho_n(x^n, \phi_n^{t, [\rho]}(x^n)) \leq d + \rho_{\max} / (qn)$. If $\rho_n(x^n, \phi_n^{t, [\rho]}(x^n)) \leq d$, then set $\phi_n(x^n, \rho) = \phi_n^{t, [\rho]}(x^n)$. Call this Case $1$. Otherwise (in Case $2$), if $d < \rho_n(x^n, \phi_n^{t, [\rho]}(x^n)) \leq d + \rho_{\max} / (qn)$, it is possible to replace exactly one symbol in the sequence $\phi_n^{t, [\rho]}(x^n)$ so that the post-corrected sequence, call it $y^n$, satisfies $\rho_n(x^n, y^n) \leq d$. Such a post-correction is possible because we have $d > 0$, the assumption in $(\ref{dist_assump})$, and 
the fact that the replacement of a symbol corresponding to maximum distortion guarantees an average distortion reduction of at least $d/n$ so that we have 
\begin{align*}
    \rho_n(x^n, y^n) &\leq \rho_n(x^n, \phi_n^{t, [\rho]}(x^n)) - \frac{d}{n}\\
    &\leq d + \frac{\rho_{\max}}{qn} - \frac{d}{n}\\
    &\leq d,
\end{align*} 
where the last inequality follows from the choice of $q = \lceil \rho_{\max} / d \rceil $. We formally write the $d$-quantizer $\phi_n$ with respect to $\rho$ as the composition of two functions, $\phi_n = w_n \circ v_n$, defined as 
\begin{align}
    &v_n(x^n, \rho) \triangleq (x^n, \rho, \phi_n^{t, [\rho]}(x^n))\\
    \notag\\
    &w_n(x^n, \rho, \phi_n^{t, [\rho]}(x^n)) \notag\\
    &\triangleq \begin{cases}
    \phi_n^{t, [\rho]}(x^n) & \text{ if } \rho_n(x^n, \phi_n^{t, [\rho]}(x^n)) \leq d  \,\,\,\,\,\, \left(\text{Case } 1 \right)\\
    y^n & \text{ if } \rho_n(x^n, \phi_n^{t, [\rho]}(x^n)) > d \,\,\,\,\,\, \left(\text{Case } 2 \right)
    \end{cases} \label{kopee}
\end{align}
where $y^n$ differs from $\phi_n^{t, [\rho]}(x^n)$ in one component and satisfies $\rho_n(x^n, y^n) \leq d$, as described above. The binary encoder $f_n$ sends 
\begin{align}
    \log \left( |\mathcal{P}_n(A)| |\mathcal{D}_n^q|  \right) + 1 &\leq \log \left( (n+1)^{J-1}  (qn+1)^{J K} \right) + 1 \notag \\
    &\leq (JK + J - 1) \log(n) + \notag\\
    &\,\,\,\,\,\,\,\,\,\,JK \log \left( \frac{\rho_{\max}}{d} + 1  \right) + JK + J 
\end{align}
bits to first identify the code $C_n^{t, [\rho]}$, followed by the binary encoding $f_n^{t, [\rho]}(\phi_n^{t, [\rho]}(x^n))$, followed by a flag bit to indicate Case $1$ vs. Case $2$ from $(\ref{kopee})$, followed by (if necessary) post-correction symbol replacement which takes at most $\log(n) + \log(K) + 2$ bits. Therefore, the expected rate of this scheme is given by
\begin{align}
    &\frac{1}{n} \mathbb{E}_p \left [ l(f_n(\phi_n(X^n, \rho))) \ln(2) \right ]\notag\\
    &\leq (JK + J - 1) \frac{\ln(n)}{n} + \frac{JK}{n} \ln \left( \frac{\rho_{\max}}{d} + 1  \right) + \frac{JK  \ln(2)}{n} \notag\\
    & \,\,\,\,\,\,\,\,\,\,+ \frac{J \ln 2}{n} + \frac{3\ln(2) + \ln(n) + \ln(K)}{n} + \notag \\
    & \,\,\,\,\,\,\,\,\,\,\,\,\,\,\,\,\,\,\,\,\,\,\,\,\,\,\,\,\,\,\frac{1}{n}\mathbb{E}_p \left [ l(f_n^{T, [\rho]}(\phi_n^{T, [\rho]}(X^n))) \ln(2) \right ] \label{typet}\\
    &= (JK + J) \frac{\ln n}{n} + \frac{W_1}{n}+ \mbox{} \notag \\
    &  + \frac{1}{n} \sum_{t \in \mathcal{P}_n(A)} p^n(T^n_A(t)) \mathbb{E}_{p} \left [  l(f_n^{t, i}(\phi_n^{t, i}(X^n))) \ln(2) | t(X^n) = t \right ] \label{conditype}\\
    &\leq (JK + J) \frac{\ln n}{n} + \frac{W_1 + \ln(2) + \epsilon}{n}+ \mbox{} \notag \\
    &  \,\,\,\,\,\,\,\,\,\,\,\,\,\,\,\,\,+ \sum_{t \in \mathcal{P}_n(A)} p^n(T^n_A(t)) \left [ \inf_{\phi_n \in \mathcal{Q}_{d, [\rho]}} \frac{ H(\nu_{p_t, \phi_n})}{n} \right ] \label{useprev}\\
    &= (JK + J) \frac{\ln n}{n} + \frac{W_1 + \ln(2) + \epsilon}{n}+ \mbox{} \notag\\
    & \,\,\,\,\,\,\,\,\,\,\,\,\,\,\,\,\,\,\,\,\,\,\,\,\,\,\,\,\,\,\,\,\,\,\,\,\,\,\,\,\,\,\,\,\,\,\,\,\,\,\,\,\,\,\,\,\,\,\,\,\,\,\,\,\,\,\,\,\mathbb{E}_p \left [  \inf_{\phi_n \in \mathcal{Q}_{d, [\rho]}} \frac{ H(\nu_{p_T, \phi_n})}{n} \right ] \notag \\
    &\leq (JK + J) \frac{\ln n}{n} + \frac{W_1 + \ln(2) + \epsilon}{n} +  \notag\\
    & \,\,\,\,\,\,\,\,\,\,\,\,\,\,\,\,\,\,\,\,\,\,\,\,\,\,\,\,\,\,\,\,\,\,\,\,\,\,\,\,\,\,\,\,\,\,\,\,\,\,\,\inf_{\phi_n \in \mathcal{Q}_{d, [\rho]}} \frac{ H( \mathbb{E}_p \left [  \nu_{p_T, \phi_n} \right])}{n} \label{concavee} \\
    &= (JK + J) \frac{\ln n}{n} + \frac{W_1 + \ln(2) + \epsilon}{n} +   \inf_{\phi_n \in \mathcal{Q}_{d, [\rho]}} \frac{ H(\nu_{p^n, \phi_n})}{n} \label{lastee}\\
    &\leq (JK + J) \frac{\ln n}{n} + \frac{W_1 + \ln(2) + \epsilon}{n} +   \inf_{\phi_n \in \mathcal{Q}_{d, \rho}} \frac{ H(\nu_{p^n, \phi_n})}{n}. \label{deflast}
\end{align}
In the last term of $(\ref{typet})$, $T = t(X^n)$ is a random type. In $(\ref{conditype})$, $W_1$ is a constant depending only on $J$, $K$, $\rho_{\max}$ and $d$. In $(\ref{useprev})$, we use $(\ref{kutee11})$ along with the fact that $X^n$ is i.i.d. according to $p$ and that conditioned on the type, $X^n$ is uniformly distributed over the type class. In $(\ref{concavee})$, we use concavity and Jensen's inequality. In $(\ref{lastee})$, we
use the same argument as in the derivation of $(\ref{vtdef})$ in the proof of Theorem \ref{silvaext}. Finally, in $(\ref{deflast})$, we use the fact that $\phi_n \in \mathcal{Q}_{d, \rho}$ implies $\phi_n \in \mathcal{Q}_{d, [\rho]}$ because of Definition \ref{defo2param}. The upper bound in $(\ref{deflast})$ holds uniformly over all $p$ and $\rho$ which enables us to write $(\ref{deflast})$ as 
\begin{align*}
    \sup_{p, \rho} \overline{R}_n(\tilde{C}_n, p, \rho) \leq (JK + J) \frac{\ln n}{n} + \frac{W_1 + \ln(2) + \epsilon}{n}.
\end{align*}
Dividing both sides by $\ln n / n$ and taking the limit establishes the result of Theorem \ref{specthm}.

\section{Proof of Theorem \ref{thmyuext} \label{thmyuextproof}}

Let $Q^{\text{NML}} \in \mathcal{P}(B^n)$ denote the normalized maximum-likelihood (NML) distribution  which is defined as
\begin{align}
    Q^{\text{NML}}(y^n) &= \frac{\sup \limits_{q \in \mathcal{P}(B)} q^n(y^n) }{S_n},
    \label{tutu}
\end{align}
where 
\begin{align}
    S_n = \sum \limits_{z^n \in B^n} \sup \limits_{p \in \mathcal{P}(B)} p^n(z^n). 
    \label{tutu2}
\end{align}
The normalization factor $S_n$ is called the Shtarkov's sum for i.i.d. distributions and $S_n$ grows only polynomially with $n$ (as can be seen from the method of types). Alternatively, Shtarkov \cite{shar1} showed the important result that $\log S_n$ is essentially (up to a discrepancy of at most $1/n$) equal to the universal lossless coding redundancy for i.i.d. source distributions.  
It is known from previous works (\cite{orlitsky1}, \cite{risannen2}, \cite{precise1}, \cite{613240}) that universal lossless coding redundancy for i.i.d. sources taking values in alphabet $B$ of size $K$ is given by   
\begin{align}
    \frac{K - 1}{2} \log (n) - \frac{K - 1}{2} \log (2 \pi) + \log \left( \frac{\Gamma\left( \frac{1}{2}\right)^K}{\Gamma \left(\frac{K}{2} \right)}\right) + o_K(1),
    \label{iidredund}
\end{align}
where $\Gamma(\cdot)$ is the gamma function and $o_K(1) \to 0$ as $n \to \infty$ at the rate determined only by $K$. Combining this with Shtarkov's result and changing base to natural log, we can express $S_n$ from $(\ref{tutu2})$ as
\begin{align}
    &S_n = \sum_{x^n \in B^n} \sup_{p \in \mathcal{P}(B)} p^n(x^n) \notag\\
    &= \exp \left( \frac{K - 1}{2} \ln n + \ln \left(  \frac{\Gamma\left( \frac{1}{2}\right)^K}{(2\pi)^{\frac{K-1}{2}} \, \Gamma \left(\frac{K}{2} \right)} \right) +  o_K(1)\, \ln (2)\right).
    \label{S_n bound }
\end{align}

Let $Z_1^n, Z_2^n, Z_3^n, \ldots $ be i.i.d. random vectors each distributed according to $Q^{\text{NML}}$. Let the random codebook $B_{\Phi_n} \subset B^n$,
\begin{align*}
    B_{\Phi_n} &= \{Z_1^n, Z_2^n, Z_3^n, \ldots \},
\end{align*}
be available to both the encoder and decoder.

Let $x^n$ be an input source sequence of type $t = t(x^n)$ and $\rho$ be the input distortion measure to the encoder. The encoder uses acceptance-rejection method (similar to \cite[Theorem 1]{https://doi.org/10.48550/arxiv.2202.04481}) to derive a subsequence $\{Z_{i_j}^n \}_{j=1}^\infty$, where ${Z}_{i_1}^n, Z_{i_2}^n, Z_{i_3}^n, \ldots $ are i.i.d. random vectors each distributed according to $(Q^{t,d,\rho})^n$. It is easy to see that
\begin{align*}
    \max_{y^n \in B^n} \frac{\prod_{i=1}^n Q^{t,d,\rho}(y_i)  }{Q^{\text{NML}}(y^n)} \leq S_n.
\end{align*}
The acceptance-rejection algorithm to construct the subsequence $\{Z_{i_j}^n\}_{j=1}^\infty$ is described as follows: 
\begin{enumerate}
    \item Set $i = 1$; $j = 1$. 
    \item Generate $U \sim \text{Unif} \left([0,1] \right)$.
    \item If $$U < \frac{ (Q^{t,d,\rho})^n(Z_i^n)}{S_n Q^{\text{NML}}(Z_{i}^n)}, \,\,\,\,\,\,\,\,\,\, (\text{success if true} )$$ then set $ i_j = i$. Set $i := i + 1$ ; $j := j + 1$. Go back to step $2$. 
    \item Else set $i := i + 1$. Go back to step $2$.
\end{enumerate}
In each iteration of the above algorithm, Step $3$ has success probability of $1/S_n$ independent of other iterations. 

Let $J(x^n)$
be the smallest integer such that $Z_{i_{J(x^n)}}^n$ satisfies 
\begin{align*}
\rho_n(x^n, Z_{i_{J(x^n)}}^n) \leq d.
\end{align*}
We set 
\begin{align}
    \Phi_n(x^n, \rho) &=  Z_{i_{J(x^n)}}^n. \label{doosrawalapan}
\end{align}
It is easy to see that $i_{J(x^n)}$ is a geometric random variable with success probability given by 
\begin{align*}
    s_{t, \rho} &= \frac{\mathbb{P}\left( \rho_n(x^n, Y^n) \leq d \right)}{S_n}
\end{align*}
so that the expected value $\mathbb{E}_c \left [ i_{J(x^n)} \right]$ is given by 
\begin{align}
    &\mathbb{E}_c \left [ i_{J(x^n)} \right] \notag \\
    &= \frac{S_n}{\mathbb{P}\left( \rho_n(x^n, Y^n) \leq d \right)}. \label{expsubskar}
\end{align}

The binary encoder $f_n$ sends $000$ if $i_{J(x^n)} = 1$, $001$ if $i_{J(x^n)} = 2$, $010$ if $i_{J(x^n)} = 3$, $011$ followed by doubly recursive Elias gamma encoding \cite{elias1} of $i_{J(x^n)}$ if $4 \leq i_{J(x^n)} \leq K^n$ and $100$ followed by fixed-rate coding of the index of $z_{i_{J}(x^n)}$ with respect to an fixed ordering of the space $B^n$ which is known to both the encoder and decoder at design time. The first three bits serve as flag bits to distinguish the cases.

Note that Elias gamma encoding of a positive integer $i$ involves writing out $N_0 = \lfloor \log i \rfloor$ zero bits followed by $\lfloor \log i \rfloor + 1$ bits for the binary representation of $i$. With one recursion, we use Elias gamma encoding to encode $N_0$, which involves writing out $N_1 = \lfloor \log N_0 \rfloor $ zero bits followed by $\lfloor \log N_0 \rfloor + 1$ bits for the binary representation of $N_0$. With a second recursion, we again use Elias gamma encoding to encode $N_1$ which involves using $2\lfloor \log N_1 \rfloor + 1 $ bits in total. Hence, to encode the integer $i$ using doubly recursive Elias encoding, the total binary length is 
\begin{align}
    &\lfloor \log i \rfloor + 1 + \lfloor \log N_0 \rfloor + 1 + 2 \lfloor \log N_1 \rfloor + 1 \label{specialelias} \\
    &\leq  \log i + \log \log i + 2 \log \log \log i + 3. \label{eliaskoinvokekar}  
\end{align}
The expression in $(\ref{specialelias})$ is undefined for $1 \leq i \leq 3$, hence the need to separately handle the case for these three values.

To finish the proof, we evaluate the expected rate of the code $\tilde{C}_n = (\Phi_n, f_n, g_n)$, where the expectation $\mathbb{E}_{c, p}[\cdot]$ is with respect to both the random code and the unknown source. Let $X^n$
be i.i.d. according to the unknown source distribution $p$ and let $\rho$ be the input distortion measure such that $(p, \rho) \in \mathcal{S}_d$. Then we have for $a = \sqrt{2 J + 2}$, 
\begin{align}
    &\frac{\ln 2}{n}\mathbb{E}_{p,c}\left[ l(f_n(\Phi_n(X^n, \rho))) \right] \notag\\
    &\stackrel{(a)}{=} \frac{\ln 2}{n} \sum_{t:||t-p||_2 \leq a \sqrt{\ln n / n}} \mathbb{E}_c\left[ p^n(T^n_A(t)) \mathbb{E}_t\left [ l(f_n(\Phi_n(X_t^n, \rho)))\right]\right] + \mbox{} \notag\\
    &\,\,\,\,\,\,\,\,\,\,\,\,\,\,\,\,\,\,\,\,\,\, \frac{\ln 2}{n} \sum_{t:||t-p||_2 > a \sqrt{\ln n / n}} \mathbb{E}_c\left [  p^n(T^n_A(t)) \mathbb{E}_t\left [ l(f_n(\Phi_n(X_t^n, \rho)))\right]\right] \notag\\
    &\stackrel{(b)}{\leq} \frac{\ln 2}{n} \sum_{t:||t-p||_2 \leq a \sqrt{\ln n / n}} p^n(T^n_A(t)) \mathbb{E}_t\left [ l(f_n(\Phi_n(X_t^n, \rho)))\right] + O\left(\frac{1}{n^2} \right) \notag\\
    &= \frac{\ln 2}{n} \sum_{t:||t-p||_2 \leq a \sqrt{\ln n / n} 
      } p^n(T^n_A(t))
     \frac{1}{|T^n_A(t)|} \sum_{x^n \in T^n_A(t)}
      \mathbb{E}_c\left [ l(f_n(\Phi_n(x^n, \rho)))\right] + O\left(\frac{1}{n^2} \right) \notag\\
    &\stackrel{(c)}{\leq} \frac{\ln 2}{n} \sum_{t:||t-p||_2 \leq a \sqrt{\ln n / n}
      } p^n(T^n_A(t)) \frac{1}{|T^n_A(t)|} \sum_{x^n \in T^n_A(t)} \mathbb{E}_c\left [ \log i_{J(x^n)} + \log \log i_{J(x^n)} + 2 \log \log \log i_{J(x^n)} | i_{J(x^n)} \geq 4   \right] + \mbox{} \notag\\
    &  \,\,\,\,\,\,\,\,\,\,\,\,\,\,\,\,\,\,\,\,\,\,O\left(\frac{1}{n} \right) \notag\\
    &\stackrel{(d)}{\leq} \frac{\ln 2}{n} \sum_{t:||t-p||_2 \leq a \sqrt{\ln n / n}
      } p^n(T^n_A(t)) \frac{1}{|T^n_A(t)|} \sum_{x^n \in T^n_A(t)} \left [ \log \mathbb{E}_c \left[i_{J(x^n)} | i_{J(x^n)} \geq 4\right] + \log \log \mathbb{E}_c\left[  i_{J(x^n)} | i_{J(x^n)} \geq 4\right] + \right. \label{elsencexp} \\
      & \left .\,\,\,\,\,\,\,\,\,\,\,\,\,\,\,\,\,\,\,\,\,\, 2 \log \log \log \mathbb{E}_c\left[ i_{J(x^n)} | i_{J(x^n)} \geq 4 \right]  \right] + O\left(\frac{1}{n} \right) \notag\\
    &\stackrel{(e)}{\leq} \frac{K + 2}{2} \frac{\ln n}{n} + O\left( \frac{\ln \ln n}{n}\right) +  \sum_{t:||t-p||_2 \leq a \sqrt{\ln n / n}
      } p^n(T^n_A(t)) R(t,d,\rho)  \label{hellom05} \\
      &\leq \mathbb{E}_p \left [ R(T,d,\rho) \right] + \frac{K + 2}{2} \frac{\ln n}{n} + O\left( \frac{\ln \ln  n}{n}\right) \notag  \\
      &= R(p,d,\rho) + \left( \frac{K}{2} + 1 \right)\frac{\ln n}{n} + o \left(\frac{\ln n}{n} \right). \label{finallybaskaryar} 
\end{align}
In equality $(a)$ above, we use the fact that conditioned on the type, $X^n$ is uniformly distributed over the type class $T^n_A(t)$, which we denote by writing $X_t^n$. In inequality $(b)$, we use Lemma \ref{lemmatypes} and the fact that the binary encoding length is always at most $n \log K + O(\log n)$, by construction. Inequality $(c)$ follows from the following manipulation: 
\begin{align}
    &\mathbb{E}_c\left [ l(f_n(\Phi_n(x^n, \rho))) \right ] \notag \\
    &\leq \mathbb{P}\left( i_{J(x^n)} \leq 3 \right) \cdot 3 + \mathbb{P}\left( i_{J(x^n)} \geq 4 \right) \mathbb{E}_c\left [ \log i_{J(x^n)} + \log \log i_{J(x^n)} + 2 \log \log \log i_{J(x^n)} + 3|i_{J(x^n)} \geq 4  \right] \notag \\
    &\leq 3 + \mathbb{E}_c\left [\log i_{J(x^n)} + \log \log i_{J(x^n)} + 2 \log \log \log i_{J(x^n)} | i_{J(x^n)} \geq 4\right]. \notag 
\end{align}
In inequality $(d)$, we use Jensen's inequality. For inequality $(e)$, we carry out the following derivation: note that there exists an $N$ depending only on $p,d,\rho$ and $\rho_{\max}$ such that for $n > N$, the result of Lemma \ref{dballlemma} applies and, from  $(\ref{S_n bound })$, $(\ref{expsubskar})$ and Lemma \ref{dballlemma}, we can write 
\begin{align*}
    &\mathbb{E}_c \left [ i_{J(x^n)} | i_{J(x^n)} \geq 4 \right]\\
    &= 4 + \mathbb{E}_c \left [ i_{J(x^n)} \right]\\
    &\leq 4 + \exp \left( n R(t,d,\rho) +  \frac{K}{2} \ln n +   \mathcal{G}_1'  \right)\\
    &\leq \exp \left( n R(t,d,\rho) +  \frac{K}{2} \ln n +   \mathcal{G}_1  \right)
\end{align*}
for some constants $\mathcal{G}_1'$ and $\mathcal{G}_1$, both also only depending on $p,d,\rho$ and $\rho_{\max}$. Hence, we can evaluate the Elias encoding expression as 
\begin{align}
    &\frac{\ln 2}{n} \left( \log \mathbb{E}_c\left[i_{J(x^n)}| i_{J(x^n)} \geq 4 \right] + \log \log \mathbb{E}_c\left[i_{J(x^n)} | i_{J(x^n)} \geq 4 \right] + 2 \log \log \log \mathbb{E}_c\left[i_{J(x^n)} | i_{J(x^n)} \geq 4\right] \right) \notag \\
    &\leq \frac{1}{n} \left(  n R(t,d,\rho) +  \frac{K}{2} \ln n +  \mathcal{G}_1 + \ln \left(\frac{1}{\ln 2} \left(  n R(t,d,\rho) +  \frac{K}{2} \ln n +  \mathcal{G}_1 \right) \right) + \frac{2}{n} \ln \log \log \mathbb{E}_c\left[ i_{J(x^n)} \right] \right) \notag \\
    &= R(t,d,\rho) + \frac{K+2}{2} \frac{\ln n}{n} + O\left( \frac{\ln \ln n}{n} \right) \label{adeelmikan}
\end{align}
where it is easy to see that the $O(\ln \ln n / n)$ term depends only on $p, d, \rho$ and $\rho_{\max}$ because $\mathcal{G}_1$ depends on the same parameters. Using $(\ref{adeelmikan})$ in $(\ref{elsencexp})$ establishes $(\ref{hellom05})$. Finally, $(\ref{finallybaskaryar})$ follows from Lemma \ref{lemmataylor} in Appendix \ref{lemmataylorproof}.

\appendices

\section{Proof of Proposition \ref{instanceofthm3from} \label{instanceofthm3fromproof}}

Fix any $d > 0$ and $(p, \rho) \in \mathcal{S}_d$. Let $\mathcal{N}_{\delta'}(p,\rho) = \{(p',\rho') \in \mathcal{P}(A) \times \mathcal{D}^{\rho_{\max}}: ||p' - p||_2 < \delta'  \text{ and }   ||\rho - \rho'||_F < \delta' \}$, for some $\delta' > 0$, be a neighborhood for which the result of Lemma \ref{dballlemma} holds. Consider a subset $\mathcal{N}_{\delta}(p, \rho)$ of this neighborhood, $\mathcal{N}_{\delta}(p, \rho) \subset \mathcal{N}_{\delta'}(p, \rho)$,  given by 
$\mathcal{N}_\delta(p,\rho) = \{(p',\rho') \in \mathcal{P}(A) \times \mathcal{D}^{\rho_{\max}}: ||p' - p||_2 < \delta  \text{ and }   ||\rho - \rho'||_F < \delta \}$, where $0 <\delta < \delta'$. Assume that $p'$ is the unknown source distribution which satisfies $||p' - p||_2 < \delta$.   

Let $Z_1^n, Z_2^n, Z_3^n, \ldots $ be i.i.d. random vectors each distributed according to $Q^{\text{NML}}$, where $Q^{\text{NML}}$ is defined in $(\ref{tutu})$. Let the random codebook $B_{\Phi_n} \subset B^n$,
\begin{align*}
    B_{\Phi_n} &= \{Z_1^n, Z_2^n, Z_3^n, \ldots \},
\end{align*}
be available to both the encoder and decoder.

We first consider only input source sequences $x^n$ with type $t = t(x^n)$ and input distortion measures $\rho'$ satisfying $(t, \rho') \in \mathcal{N}_{\delta'}(p, \rho)$.
The encoder uses acceptance-rejection method (similar to the proof of Theorem \ref{thmyuext}) to derive a subsequence $\{Z_{i_j}^n \}_{j=1}^\infty$, where ${Z}_{i_1}^n, Z_{i_2}^n, Z_{i_3}^n, \ldots $ are i.i.d. random vectors each distributed according to $(Q^{t,d,\rho'})^n$. 

Let $J(x^n)$
be the smallest integer such that $Z_{i_{J(x^n)}}^n$ satisfies 
\begin{align*}
\rho'_n(x^n, Z_{i_{J(x^n)}}^n) \leq d.
\end{align*}
It is easy to see that $i_{J(x^n)}$ is a geometric random variable with success probability given by 
\begin{align*}
    s_{t, \rho'} &= \frac{\mathbb{P}\left( \rho'_n(x^n, Y^n) \leq d \right)}{S_n},
\end{align*}
where $S_n$ is defined in $(\ref{tutu2})$. The expected value $\mathbb{E}_c \left [ i_{J(x^n)} \right]$ is given by 
\begin{align}
    &\mathbb{E}\left [ i_{J(x^n)} \right ] \notag \\
    &= \frac{S_n}{\mathbb{P}\left(\rho'_n(x^n, Y^n) \leq d \right)}\notag \\
    &\leq \exp \left( n R(t,d,\rho') +  \frac{K}{2} \ln n + O(1)\right), \label{pakisadeel}
\end{align}
where the $O(1)$ term depends only on $p, d,\rho, \rho_{\max}$ and the alphabet sizes\footnote{Since the alphabet sizes are fixed throughout the paper, we ignore the dependence on them from now on.}, which is easy to see from $(\ref{S_n bound })$ and the statement of Lemma \ref{dballlemma}. It turns out that the upper bound in $(\ref{pakisadeel})$ not only holds in expectation but also (up to a $\ln \ln n$ factor) holds with high probability, as we will show next. This property will be crucial in showing the existence of a deterministic codebook. 

Let 
\begin{align*}
    \gamma_n = 1 + \frac{\ln (J^2 K^2 + J - 1)}{\ln \ln n}.
\end{align*}
Denoting the probability law associated with the random codebook $B_{\Phi_n}$ by $\mathbb{P}_c(\cdot)$, we have  
\begin{align}
    &\mathbb{P}_c\left( i_{J(x^n)} > \exp \left( n R(t,d,\rho') +  \frac{K}{2} \ln n + \gamma_n \ln \ln n +  O(1)\right) \right) \notag \\
    &\leq (1-s_{t,\rho'})^{\exp \left( n R(t,d,\rho') +  \frac{K}{2} \ln n + \gamma_n \ln \ln n +  O(1)\right) - 1}\notag \\
    &\leq \left (1- \exp\left(-n R(t,d,\rho') - \frac{K}{2} \ln n - O(1) \right) \right )^{\exp \left( n R(t,d,\rho') +  \frac{K}{2} \ln n + \gamma_n \ln \ln n +  O(1)\right) - 1} \label{bigOcancel} \\
    &\leq \exp\left(- \exp\left(\gamma_n \ln \ln n \right) + \exp\left(-n R(t,d,\rho') - \frac{K}{2} \ln n - O(1) \right)  \right) \notag \\
    &= O\left( \frac{1}{n^{J^2 K^2 + J - 1}} \right), \label{comeononon}
\end{align}
where the big O term in the last equality above again depends only on $p, d, \rho$ and $\rho_{\max}$. Also note that the two "$O(1)$" terms appearing in $(\ref{bigOcancel})$ are identical which explains the cancellation occurring in the next inequality.

The bound in $(\ref{comeononon})$ holds for a particular $x^n \in T^n_A(t)$. Now if we let $X_t^n  \sim \text{Unif}(T^n_A(t))$ be a random sequence uniformly distributed over the type class $T^n_A(t)$, then it is easy to see from $(\ref{comeononon})$ that we have 
\begin{align}
    &\mathbb{P}_{t,c}\left( i_{J(X_t^n)} > \exp \left( n R(t,d,\rho') +  \frac{K}{2} \ln n + \gamma_n \ln \ln n +  O(1)\right) \right) \notag \\
    &= O\left( \frac{1}{n^{J^2 K^2 + J - 1}} \right). \label{comeononon2}
\end{align}
We used $\mathbb{P}_{t,c}$ above to denote the probability law associated with the random sequence $X_t^n \sim \text{Unif}(T^n_A(t))$ and the random codebook. Note that $(\ref{comeononon2})$ holds for an arbitrary input $(t,\rho' ) \in \mathcal{N}_{\delta'}(p, \rho)$ to the encoder. But we want that with high probability, the integer index is uniformly "small" over the entire set $\mathcal{N}_{\delta'}(p, \rho)$. For this, we use a straightforward corollary of Proposition \ref{propo2}: for a given blocklength $n$ and distortion level $d$, there are $ \xi \leq (n+1)^{J^2 K^2 - 1} + 1$ equivalence classes of $\mathcal{D}^{\rho_{\max}}$, denoted by $[\mathcal{D}]_{\rho_1}$, $[\mathcal{D}]_{\rho_2}$, \ldots, $[\mathcal{D}]_{\rho_{\xi}}$, where $\rho_1, \rho_2, \ldots, \rho_\xi$ are arbitrarily chosen representative distortion measures. A $d$-semifaithful code with respect to a distortion measure $\rho$ is also $d$-semifaithful with respect to all distortion measures $\rho' \in [\mathcal{D}]_{\rho}$ in the same equivalence class. We will make the choice of representative distortion measures $\rho_1, \rho_2, \ldots, \rho_{\xi} $ be a function of the type $t$. For every $n$-type $t$ and every equivalence class $[\mathcal{D}]_{\rho_i}$, we can choose the representative distortion measure $\rho_i^t \in [\mathcal{D}]_{\rho_i}$ to satisfy 
\begin{align}
    R(t,d,\rho_i^t) \leq \inf_{\tilde{\rho} \in [\mathcal{D}]_{ \rho_i  }} R(t,d,\tilde{\rho}) + \epsilon  \label{phoit}
\end{align}
for any $\epsilon > 0$. Henceforth, we will choose $\epsilon = 1/n$ and the representative distortion measures, chosen differently for each type, will be denoted by $\rho_1^t, \rho_2^t, \ldots, \rho^t_{\xi}$.  

We next use subscript "$\,T\,$" to denote the probability law associated with the collection of random sequences $\{X_t^n : t \in \mathcal{P}_n(A) \}$. 
Taking a union bound over all types and equivalence classes of distortion measures in $\mathcal{N}_{\delta'}(p, \rho)$ gives us that 
\begin{align*}
    &\mathbb{P}_{T,c} \left( \bigcup_{t :||t-p||_2 < \delta'} \bigcup_{\substack{\tilde{\rho} \in \{\rho_1^t, \ldots, \rho_{\xi}^t \}\\ ||\tilde{\rho} - \rho||_F < \delta' } } \left \{  i_{J(X_t^n)} > \exp \left( n R(t,d,\tilde{\rho}) +  \frac{K}{2} \ln n + \gamma_n \ln \ln n +  O(1)\right) \right \}  \right)\\
    &= O\left(\frac{(n+1)^{J-1} (n+1)^{J^2 K^2  - 1}}{n^{J^2 K^2 + J - 1}} \right)\\
    &= O\left(\frac{1}{n} \right).
\end{align*}
Also note that 
\begin{align}
    &\mathbb{P}_{T,c} \left( \bigcup_{t :||t-p||_2 < \delta'} \bigcup_{\substack{\tilde{\rho} \in \{\rho_1^t, \ldots, \rho_{\xi}^t \}\\ ||\tilde{\rho} - \rho||_F < \delta' } } \left \{  i_{J(X_t^n)} > \exp \left( n R(t,d,\tilde{\rho}) +  \frac{K}{2} \ln n + \gamma_n \ln \ln n +  O(1)\right) \right \}  \right) \notag \\
    &= \mathbb{E}_{T,c} \left [ \mathds{1} \left(  \bigcup_{t :||t-p||_2 < \delta'} \bigcup_{\substack{\tilde{\rho} \in \{\rho_1^t, \ldots, \rho_{\xi}^t \}\\ ||\tilde{\rho} - \rho||_F < \delta' } } \left \{  i_{J(X_t^n)} > \exp \left( n R(t,d,\tilde{\rho}) +  \frac{K}{2} \ln n + \gamma_n \ln \ln n +  O(1)\right) \right \} \right) \right ] \notag \\
    &= \mathbb{E}_c \left [ \mathbb{E}_T \left [  \mathds{1} \left(  \bigcup_{t :||t-p||_2 < \delta'} \bigcup_{\substack{\tilde{\rho} \in \{\rho_1^t, \ldots, \rho_{\xi}^t \}\\ ||\tilde{\rho} - \rho||_F < \delta' } } \left \{  i_{J(X_t^n)} > \exp \left( n R(t,d,\tilde{\rho}) +  \frac{K}{2} \ln n + \gamma_n \ln \ln n +  O(1)\right) \right \} \right) \Bigg | B_{\Phi_n} \right ]  \right] \label{explkar1}\\
    &= O\left(\frac{1}{n} \right). \notag
\end{align}
The above result implies that there exists a deterministic codebook,  call it $B_{\phi_n}$, such that 
\begin{align}
     &\mathbb{E}_T \left [  \mathds{1} \left(  \bigcup_{t :||t-p||_2 < \delta'} \bigcup_{\substack{\tilde{\rho} \in \{\rho_1^t, \ldots, \rho_{\xi}^t \}\\ ||\tilde{\rho} - \rho||_F < \delta' } } \left \{  i_{J(X_t^n)} > \exp \left( n R(t,d,\tilde{\rho}) +  \frac{K}{2} \ln n + \gamma_n \ln \ln n +  O(1)\right) \right \} \right) \Bigg | B_{\Phi_n} = B_{\phi_n} \right ] \label{explkar2}\\
     &= O\left(\frac{1}{n} \right). \notag 
\end{align}
In $(\ref{explkar1})$, $i_{J(X_t^n)}$ is a random variable whose randomness stems from both the random codebook $B_{\Phi_n}$ and the random sequence $X_t^n$, whereas in $(\ref{explkar2})$, the randomness of $i_{J(X_t^n)}$ only stems from the random sequence $X^n_t$.

The result in $(\ref{explkar2})$ implies that 
\begin{align}
    &\mathbb{P}_T\left(   \bigcup_{t :||t-p||_2 < \delta'} \bigcup_{\substack{\tilde{\rho} \in \{\rho_1^t, \ldots, \rho_{\xi}^t \}\\ ||\tilde{\rho} - \rho||_F < \delta' } } \left \{  i_{J(X_t^n)} > \exp \left( n R(t,d,\tilde{\rho}) +  \frac{K}{2} \ln n + \gamma_n \ln \ln n +  O(1)\right) \right \} \Bigg | B_{\Phi_n} = B_{\phi_n} \right) \notag \\
    &\stackrel{(a)}{=} \mathbb{P}_T\left(   \bigcup_{t :||t-p||_2 < \delta'} \bigcup_{\substack{\tilde{\rho} \in \{\rho_1^t, \ldots, \rho_{\xi}^t \}\\ ||\tilde{\rho} - \rho||_F < \delta' } } \left \{  i_{J(X_t^n)} > \exp \left( n R(t,d,\tilde{\rho}) +  \frac{K}{2} \ln n + \gamma_n \ln \ln n +  O(1)\right) \right \} \right) \notag  \\
    &= O\left(\frac{1}{n} \right), \label{nowmakethisprecise}
\end{align}
where equality $(a)$ above follows from the independence of the random codebook $B_{\Phi_n}$ and the random source sequence $X_t^n$. Now we have a deterministic codebook $B_{\phi_n}$ which, with high probability, has uniformly good performance (i.e., small value of index $i_{J(X_t^n)}$) in encoding a random sequence $X_t^n \sim \text{Unif}(T^n_A(t))$ for any type $t$ and any of the chosen representative distortion measures $ \tilde{\rho} \in \{\rho_1^t, \rho_2^t, \ldots, \rho_\xi^t\}$, such that $(t,\tilde{\rho}) \in \mathcal{N}_{\delta'}(p,\rho)$.

In the result of $(\ref{nowmakethisprecise})$, both the $O(1)$ and $O(1/n)$ terms depend only on $p, d, \rho $ and $\rho_{\max}$. This means that exist some numbers $N$, $\mathcal{G}_1$ and $\mathcal{G}_2$ depending only on $p, d, \rho$ and $\rho_{\max}$ such that for $n > N$, we have 
\begin{align}
    &\mathbb{P}_T\left(   \bigcup_{t :||t-p||_2 < \delta'} \bigcup_{\substack{\tilde{\rho} \in \{\rho_1^t, \ldots, \rho_{\xi}^t \}\\ ||\tilde{\rho} - \rho||_F < \delta' } } \left \{  i_{J(X_t^n)} > \exp \left( n R(t,d,\tilde{\rho}) +  \frac{K}{2} \ln n + \gamma_n \ln \ln n +  \mathcal{G}_1\right) \right \} \right)\notag \\
    &\leq \frac{\mathcal{G}_2}{n}. \label{propertyofone} 
\end{align}
Now we construct the universal distortion code $\tilde{C}_n = (\phi_n, g_n, f_n)$ for $n > N$, with codebook $B_{\phi_n}$, where $B_{\phi_n}$ satisfies $(\ref{propertyofone})$. Let $B_{\phi_n} = \{z_1^n, z_2^n, \dots \}$. Given any input source sequence $x^n$ with arbitrary type $t = t(x^n)$ and input distortion measure $\rho'$ satisfying $||\rho' - \rho||_F < \delta$, let $\rho' \in [\mathcal{D}]_{\rho_i^t}$ for some $1 \leq i \leq \xi$, define \begin{align*}
    \kappa(t,\rho') \triangleq \exp \left( n R(t,d,\rho_i^t) +  \frac{K}{2} \ln n + \gamma_n \ln \ln n +  \mathcal{G}_1 \right), 
\end{align*}
let 
\begin{align*}
    i_{J(x^n)} =  \min_{i: \rho'_n(x^n, z_i^n) \leq d} i 
\end{align*}
and set $\phi_n(x^n, \rho') = z_{i_{J(x^n)}}^n$.

The binary encoder $f_n$ sends $000$ if $i_{J(x^n)} = 1$, $001$ if $i_{J(x^n)} = 2$, $010$ if $i_{J(x^n)} = 3$, $011$ followed by doubly recursive Elias gamma encoding\footnote{The doubly recursive Elias gamma encoding is described in the proof of Theorem \ref{thmyuext}, see $(\ref{specialelias})$ and $(\ref{eliaskoinvokekar})$} of $i_{J(x^n)}$ if $4 \leq i_{J(x^n)} \leq K^n$ and $100$ followed by fixed-rate coding of the index of $z_{i_{J}(x^n)}$ with respect to a fixed ordering of the space $B^n$ which is known to both the encoder and decoder at design time. The first three bits serve as flag bits to distinguish the cases.

To finish the proof, we evaluate the expected rate of the code $\tilde{C}_n$. Let $X^n$
be i.i.d. according to the unknown source distribution $p'$ where $||p' - p||_2 < \delta$. Then for any input distortion measure $\rho'$ satisfying $||\rho' - \rho||_F < \delta$ with $\rho' \in [\mathcal{D}]_{\rho_i^t}$ for some $1 \leq i \leq \xi$, we have for $a = \sqrt{2 J + 2}$,
\begin{align}
    &\frac{\ln 2}{n}\mathbb{E}_{p'}\left[ l(f_n(\phi_n(X^n, \rho'))) \right] \notag\\
    &\stackrel{(a)}{=} \frac{\ln 2}{n} \sum_{t:||t-p'||_2 \leq a \sqrt{\ln n / n}} p'^n(T^n_A(t)) \mathbb{E}_t\left [ l(f_n(\phi_n(X_t^n, \rho')))\right] + \mbox{} \notag\\
    &\,\,\,\,\,\,\,\,\,\,\,\,\,\,\,\, \frac{\ln 2}{n} \sum_{t:||t-p'||_2 > a \sqrt{\ln n / n}} p'^n(T^n_A(t)) \mathbb{E}_t\left [ l(f_n(\phi_n(X_t^n, \rho')))\right] \notag\\
    &\stackrel{(b)}{\leq} \frac{\ln 2}{n} \sum_{t:||t-p'||_2 \leq a \sqrt{\ln n / n}} p'^n(T^n_A(t)) \mathbb{E}_t\left [ l(f_n(\phi_n(X_t^n, \rho')))\right] +  
     O\left(\frac{1}{n^2} \right) \notag\\
    &= \frac{\ln 2}{n} \sum_{t:||t-p'||_2 \leq a \sqrt{\ln n / n} 
      } p'^n(T^n_A(t)) \mathbb{P}\left( i_{J(X^n_t)} \leq \kappa(t,\rho') \right)\mathbb{E}_t\left [ l(f_n(\phi_n(X_t^n, \rho'))) | i_{J(X^n_t)} \leq \kappa(t,\rho')\right] + \mbox{} \notag\\
    &\,\,\,\,\,\,\,\,\,\,\,\,\,\,\,\, \frac{\ln 2}{n} \sum_{t:||t-p'||_2 \leq a \sqrt{\ln n / n} } p'^n(T^n_A(t)) \mathbb{P}\left(i_{J(X^n_t)} > \kappa(t,\rho') \right) \mathbb{E}_t\left [ l(f_n(\phi_n(X_t^n, \rho'))) | i_{J(X^n_t)} > \kappa(t,\rho')\right] +   O\left(\frac{1}{n^2} \right) \notag\\
    &\stackrel{(c)}{\leq} \frac{\ln 2}{n} \sum_{t:||t-p'||_2 \leq a \sqrt{\ln n / n} } p'^n(T^n_A(t)) \mathbb{E}_t\left [ l(f_n(\phi_n(X_t^n, \rho'))) | i_{J(X^n_t)} \leq \kappa(t,\rho')\right] +  O\left(\frac{1}{n} \right) \notag\\
    &\stackrel{(d)}{\leq} \frac{\ln 2}{n} \sum_{t:||t-p'||_2 \leq a \sqrt{\ln n / n} } p'^n(T^n_A(t)) \left(\log \kappa(t,\rho') + \log \log \kappa(t,\rho') + 2 \log \log \log \kappa(t,\rho') \right) +     O\left(\frac{1}{n} \right) \notag \\
    &\stackrel{(e)}{=}   \sum_{t:||t-p'||_2 \leq a \sqrt{\ln n / n} } p'^n(T^n_A(t)) R(t,d,\rho_i^t)  + \frac{K+ 2}{2} \frac{\ln n}{n} + O \left(\frac{\ln \ln n}{n} \right) \notag \\
    &\stackrel{(f)}{\leq} \mathbb{E}_{p'} \left [ R(T, d, \rho') \right ]  +  \frac{K+ 2}{2} \frac{\ln n}{n} + O \left(\frac{\ln \ln n}{n} \right). \label{finallyunif61}
\end{align}
In all of the above, we assume sufficiently large $n$ so that $(t, \rho') \in \mathcal{N}_{\delta'}(p, \rho)$ whenever $||t-p'|| \leq a \sqrt{\ln n / n}$. In equality $(a)$ above, we use the fact that conditioned on the type, $X^n$ is uniformly distributed over the type class $T^n_A(t)$, which we denote by writing $X_t^n$. In inequality $(b)$, we use Lemma \ref{lemmatypes} and the fact that the binary encoding length is always at most $n \log K + O(\log n)$, by construction. In inequality $(c)$, we use the fact that the codebook $B_{\phi_n}$   
satisfies $(\ref{propertyofone})$ and that the binary encoding length is always at most $n \log K + O(\log n)$. In inequality $(d)$, we upper bound the binary encoding length by the Elias gamma encoding of $\kappa(t, \rho')$.  In inequality $(e)$, we evaluated the Elias encoding expression as 
\begin{align*}
    &\frac{\ln 2}{n} \left( \log \kappa(t, \rho') + \log \log \kappa(t, \rho') + 2 \log \log \log \kappa(t, \rho') \right)\\
    &= \frac{1}{n} \left(  n R(t,d,\rho_i^t) +  \frac{K}{2} \ln n + \gamma_n \ln \ln n +  \mathcal{G}_1 + \ln \left(\frac{1}{\ln 2} \left(  n R(t,d,\rho_i^t) +  \frac{K}{2} \ln n + \gamma_n \ln \ln n +  \mathcal{G}_1 \right) \right) + \mbox{} \right . \\
    &\left . \,\,\,\,\,\,\,\,\,\,\,\,\,\,\,\,\,\,\,\,\,\,\,\,\,\,\,\,\,\,\,\,\,\,\,\,\,\,\,\,\,\, \frac{2}{n} \ln \log \log \kappa(t, \rho') \right)\\
    &= R(t,d,\rho_i^t) + \frac{K+2}{2} \frac{\ln n}{n} + O\left( \frac{\ln \ln n}{n} \right),
\end{align*}
where it is easy to see that $O(\ln \ln n / n)$ depends only on $p, d, \rho$ and $\rho_{\max}$ because $\mathcal{G}_1$ depends only on $p, d, \rho$ and $\rho_{\max}$, and we can use $R(t,d,\rho_i^t) \leq \ln K$. Inequality $(f)$ follows from the way we chose the representative distortion measure in $(\ref{phoit})$. Since the upper bound in $(\ref{finallyunif61})$ holds uniformly for all $(p', \rho') \in \mathcal{N}_{\delta}(p, \rho)$, we can say that there exist positive $N$ and $\mathcal{F}$ depending only on $\mathcal{N}_{\delta}(p, \rho)$ (and $d$ and $\rho_{\max}$ of course) such that for $n > N$, 
\begin{align*}
     &\frac{\ln 2}{n} \mathbb{E}_{p'} \left [ l(f_n(\phi_n(X^n, \rho'))) \right ]\\
     &\leq \mathbb{E}_{p'} \left [ R(T, d, \rho') \right ] + \frac{K + 2}{2} \frac{\ln n}{n} + \mathcal{F}\frac{\ln \ln n}{n}. 
\end{align*} 
This finishes the proof of Proposition \ref{instanceofthm3from}.

\section{Proof of Lemma \ref{contlemma} \label{contlemmaproof}}

Let $\{(p_n, d_n, \rho_n) \}_{n=1}^\infty$ be a sequence of triples converging to $(p, d, \rho)$. Let $W^*_{B|A}[p_n, d_n, \rho_n]$ be any minimizer corresponding to $(p_n, d_n, \rho_n)$. Let $W^*_{B|A}[p_\infty, d_\infty, \rho_\infty]$ be any subsequential limit of $W^*_{B|A}[p_n, d_n, \rho_n]$ with respect to the $||\cdot||_F$ metric as $n$ goes to infinity. Mathematically, this implies that there exists a subsequence $\{n_l\}$ such that 
\begin{align*}
    \lim_{l \to \infty} ||W^*_{B|A}[p_\infty, d_\infty, \rho_\infty] - W^*_{B|A}[p_{n_l}, d_{n_l}, \rho_{n_l}]||_F = 0.
\end{align*}
It suffices to show that $W^*_{B|A}[p_\infty, d_\infty, \rho_\infty] = W^*_{B|A}[p,d,\rho]$.

Clearly, we have 
\begin{align}
    \sum_{j \in A} \sum_{k \in B} p_{n_l}(j) W^*_{B|A}[p_{n_l}, d_{n_l}, \rho_{n_l}](k|j) \rho_{n_l}(j, k) - d_{n_l} &\leq 0. \label{limlen}
\end{align}
Taking the limit as $l$ goes to infinity in $(\ref{limlen})$ gives 
\begin{align*}
     \sum_{j \in A, k \in B} p(j) W^*_{B|A}[p_\infty, d_\infty, \rho_\infty](k|j) \rho(j, k) - d  &\leq 0,
\end{align*}
which shows that the subsequential limit $W^*_{B|A}[p_\infty, d_\infty, \rho_\infty]$ is feasible for the given $(p,d,\rho)$. We already know from the optimality of $W^*_{B|A}[p,d,\rho]$ that  
\begin{align}
    I\left(p, W^*_{B|A}[p,d,\rho] \right) \leq I \left(p, W^*_{B|A}[p_\infty, d_\infty, \rho_\infty] \right). \label{juv2}
\end{align}
Let $\tilde{d}_n$ be the distortion induced by the joint distribution $( p_n \times W^*_{B|A}[p,d,\rho])$ with respect to distortion measure $\rho_n$. Then we have 
\begin{align}
    I\left ( p_n, W^*_{B|A}[p,d,\rho] \right) \geq R(p_n, \tilde{d}_n, \rho_n). \label{carol1}
\end{align}
By the convexity of $R(p,d,\rho)$ in $d$, $(\ref{carol1})$ implies (e.g.,~\cite[Lemma~5.16]{Royden:Real})
\begin{align}
& I\left ( p_n, W^*_{B|A}[p,d,\rho] \right) \\
&\geq R(p_n, \tilde{d}_n, \rho_n) \notag \\
&\geq R(p_n, d_n, \rho_n) + \mbox{ } \notag \\
& \phantom{= I \Bigg(} \frac{\partial R(p_n, d_n, \rho_n)}{\partial d} (\tilde{d}_n - d_n) \notag \\
    & = I\left(p_n, W^*_{B|A}[p_n, d_n, \rho_n] \right) + \mbox{ } \notag \\
    & \phantom{= I \Bigg(} \frac{\partial R(p_n, d_n, \rho_n)}{\partial d} (\tilde{d}_n - d_n). \label{contlemmahuy}
\end{align}
Since $d > 0$, there exists an $N > 0$ and $\epsilon > 0$ such that $d_n \geq \epsilon$ for all $n \geq N$. From the assumption in $(\ref{dist_assump})$ and from the convexity of $R(p,d,\rho)$ in $d$, we have (i) $R(p,d,\rho) \leq \ln(K)$ and (ii) $  | \partial R(p_n,d_n,\rho_n)/\partial d | \leq \ln(K)/ \epsilon$. Furthermore, since $\tilde{d}_n$ is continuous as a function of $p_n$ and $\rho_n$, we have that $\tilde{d}_n$ tends to $d$ in the limit as $n$ goes to infinity. Hence, writing $(\ref{contlemmahuy})$ using the subsequence $\{n_l\}$, we have 
\begin{align}
    I\left ( p_{n_l}, W^*_{B|A}[p,d,\rho] \right) &\geq I\left(p_{n_l}, W^*_{B|A}[p_{n_l}, d_{n_l}, \rho_{n_l}] \right) - \notag \\
    & \,\,\,\,\,\,\,\,\,\,\,\,\,\,\,\,\,\,\,\,\,\,\frac{\partial R(p_{n_l}, d_{n_l}, \rho_{n_l})}{\partial d} (\tilde{d}_{n_l} - d_{n_l}) \label{kutops}
\end{align}
for sufficiently large $l$. Now taking the limit as $l$ goes to infinity in $(\ref{kutops})$, we have 
\begin{align}
    \lim_{l \to \infty} I ( p_{n_l}, W^*_{B|A}[p,d,\rho] ) &\geq \lim_{l \to \infty} I(p_{n_l}, W^*_{B|A}[p_{n_l}, d_{n_l}, \rho_{n_l}] ) \notag \\
    I(p,W^*_{B|A}[p,d,\rho]) &\geq I(p, W^*_{B|A}[p_\infty,d_\infty,\rho_\infty]), \label{juv}
\end{align}
where the last inequality follows by continuity of mutual information $I(p, W)$ as a function of the joint $p \times W$. Since $W^*_{B|A}[p,d,\rho]$ is unique, it follows from $(\ref{juv2})$ and $(\ref{juv})$ that $W^*_{B|A}[p,d,\rho] = W^*_{B|A}[p_\infty, d_\infty, \rho_\infty]$.

\section{Convergence of $\mathbb{E}_p\left [R(T,d,\rho) \right]$ to $R(p,d,\rho)$ \label{lemmataylorproof}}

\begin{lemma}
Fix $d > 0$ and any $(p,\rho) \in \mathcal{S}_d$, where $\mathcal{S}_d$ is defined in Definition \ref{defineSdset}. Then we have  
\begin{align*}
    \mathbb{E}_p \left [ R(T,d,\rho) \right] - R(p,d,\rho)   \leq o\left(\frac{\ln n}{n} \right).
\end{align*}
\label{lemmataylor}
\end{lemma}
\begin{IEEEproof}
Similar to $(\ref{altrd})$,
the rate-distortion function has a characterization in terms of the lower mutual information introduced in \cite[(23)]{yang1}, 
\begin{align*}
    R(p,d,\rho) &= \inf_{q \in \mathcal{P}(B)} I_l(q, p,d,\rho).
\end{align*}
The lower mutual information is defined as 
\begin{align*}
    I_l(q, p, d, \rho) \triangleq H(p) + H(q) - \sup_{s \in \mathcal{S}(q, p, d, \rho) } H(s),
\end{align*}
where $\mathcal{S}(q, p, d, \rho) \subset \mathcal{P}(A \times B)$ is the set of all joint distributions $s$ with marginals $p$ and $q$ on alphabets $A$ and $B$, respectively, such that $\mathbb{E}[\rho(X, Y)] \leq d$ for $(X, Y) \sim s$. Properties of $I_l(q,p,d,\rho)$ can be found in \cite[Lemmas 1 and 2]{yang1}. In particular, it follows from \cite[Lemma 2]{yang1} that for any fixed $p,d$ and $\rho$,  $I_l(Q^{p,d,\rho},p',d,\rho)$ is second-order differentiable in its second argument for any $p'$ satisfying $||p' - p|| \leq \delta$ for some $\delta > 0$.   

For $a = \sqrt{2 J +2}$, we have 
\begin{align}
    &\mathbb{E}_p [R(T,d,\rho)] \notag\\
    &= \sum_{t \in \mathcal{P}_n(A)} p^n(T^n_A(t)) R(t,d,\rho) \notag\\
    &= \sum_{t:||t - p||_2 \leq a \sqrt{\ln n / n}}  p^n(T^n_A(t)) R(t,d,\rho) \notag\\
    & \,\,\,\,\,\,\,\,\,\,\,\,+ \sum_{t:||t - p||_2 > a \sqrt{\ln n / n}}  p^n(T^n_A(t)) R(t,d,\rho) \notag \\
    &\stackrel{(a)}{\leq} \sum_{t:||t - p||_2 \leq a \sqrt{\ln n / n}}  p^n(T^n_A(t)) R(t,d,\rho) +   \ln(K) \frac{e^{J-1}}{n^2} \notag \\
    &= \sum_{t:||t - p||_2 \leq a \sqrt{\ln n / n}}  p^n(T^n_A(t)) \left[ \inf_{q \in \mathcal{P}(B)} I_l(q,t,d,\rho)\right] +   \ln(K) \frac{e^{J-1}}{n^2} \notag \\
    &\leq \sum_{t:||t - p||_2 \leq a \sqrt{\ln n / n}}  p^n(T^n_A(t)) \left[  I_l(Q^{p,d,\rho},t,d,\rho)\right] +   \ln(K) \frac{e^{J-1}}{n^2} \notag \\
    &\stackrel{(b)}{=} I_l(Q^{p,d,\rho},p,d,\rho) +  \sum_{t:||t - p||_2 \leq a \sqrt{\ln n / n}}  p^n(T^n_A(t)) \notag \\
    & \,\,\left(  \left \langle \frac{\partial I_l(Q^{p,d,\rho},p',d,\rho)}{\partial p'} \Bigg |_{p' = p}, t - p \right \rangle +  \frac{1}{2} (t-p)' \frac{\partial^2 I_l(Q^{p,d,\rho},p',d,\rho)}{\partial p'^2} \Bigg |_{p' = p} (t-p) \right . \notag \\
    & \,\,\,\,\,\,\,\,\,\,\,\,\,\,\,\,\,\,\,\,\,\,\,\,\,\,\,  + \,  o(||t-p||^2)  \Big ) +    \ln(K) \frac{e^{J-1}}{n^2} \notag\\
    &= R(p,d,\rho) + o\left(\frac{\ln n}{n} \right) \notag \\
    &+ \sum_{t:||t - p||_2 \leq a \sqrt{\ln n / n}}  p^n(T^n_A(t))   \left \langle \frac{\partial I_l(Q^{p,d,\rho},p',d,\rho)}{\partial p'} \Bigg |_{p' = p}, t - p \right \rangle  + \notag\\ 
    & \,\,\,\,\, \frac{1}{2} \sum_{t:||t - p||_2 \leq a \sqrt{\ln n / n}}  p^n(T^n_A(t)) (t-p)' \frac{\partial^2 I_l(Q^{p,d,\rho},p',d,\rho)}{\partial p'^2} \Bigg |_{p' = p} (t-p).  \label{subeeresult2}
 \end{align}
Inequality $(a)$ uses Lemma \ref{lemmatypes} and the fact that $R(t,d,\rho)  \leq \ln(K)$, which follows from $(\ref{dist_assump})$. In equality $(b)$, we assume $n$ large enough so that $t$ satisfies $||t - p||_2 \leq \delta$ which allows us to use the second-order differentiability property as mentioned in the beginning of the proof. Also, equality $(b)$ uses a slightly lesser known form of Taylor's Theorem \cite[p. 290]{hardy_korner_2008}. We now show that the last two terms in $(\ref{subeeresult2})$ are $O(1/n)$. Since we have 
\begin{align*}
    \sum_{t \in \mathcal{P}_n(A)} p^n(T^n_A(t)) (t(j) - p(j)) = 0
\end{align*}
for all $j \in A$, it follows (similar to the approach used in \cite[Theorem 2]{yu1}) that
\begin{align*}
    &\Bigg | \sum_{t: ||t-p||_2 \leq a \sqrt{\ln n / n}} p^n(T^n_A(t)) (t(j) - p(j)) \Bigg | \\
    &= \Bigg | \sum_{t: ||t-p||_2 > a \sqrt{\ln n / n}} p^n(T^n_A(t)) (t(j) - p(j)) \Bigg |\\
    &\leq \sum_{t: ||t-p||_2 > a \sqrt{\ln n / n}} p^n(T^n_A(t))\\
    &\leq \frac{e^{J-1}}{n^2} \,\,\,\,\,\,\,\,\,\,\,\,\,\, \left(\text{from Lemma \ref{lemmatypes}} \right)
\end{align*}
and therefore,
\begin{align}
    &\Bigg |\sum_{t:||t - p||_2 \leq a \sqrt{\ln n / n}}  p^n(T^n_A(t))  \left \langle \frac{\partial I_l(Q^{p,d,\rho},p',d,\rho)}{\partial p'} \Bigg |_{p' = p}, t - p \right \rangle \Bigg | \notag \\
    &= \Bigg | \sum_{t:||t - p||_2 \leq a \sqrt{\ln n / n}}  p^n(T^n_A(t)) \sum_{j=1}^J   \frac{\partial I_l(Q^{p,d,\rho},p',d,\rho)}{\partial p'_j} \Bigg |_{p'_j = p_j} ( t(j) - p(j) ) \Bigg | \notag \\
    &\leq \sum_{j=1}^J \Bigg |\frac{\partial I_l(Q^{p,d,\rho},p',d,\rho)}{\partial p'_j} \Bigg |_{p'_j = p_j}  \Bigg | \cdot \Bigg |\sum_{t:||t - p||_2 \leq a \sqrt{\ln n / n}}  p^n(T^n_A(t)) (t(j) - p(j)) \Bigg| \notag \\
    &\leq \sum_{j=1}^J \Bigg |\frac{\partial I_l(Q^{p,d,\rho},p',d,\rho)}{\partial p'_j} \Bigg |_{p'_j = p_j}  \frac{e^{J-1}}{n^2} \notag \\
    &= O\left(\frac{1}{n^2} \right). \label{resulta1niya}
\end{align}
For the second term in $(\ref{subeeresult2})$, we can write it as  
\begin{align}
    &\frac{1}{2}\sum_{t:||t-p||_2\leq a \sqrt{\ln n / n}} p^n(T^n_A(t))  \sum_{i=1}^J \sum_{j=1}^J  (t(i)-p(i)) \notag \\
    &\,\,\,\,\,\,\,\,\,\,\,\,\,\,\,\,\,\,\,\,\,\,\,\,\,\,\,\,\,\,\,\,\,\,\,\,\,\frac{\partial^2 I_l(Q^{p,d,\rho},p',d,\rho)}{\partial p'_i \partial p'_j} \Bigg |_{p'_i=p_i,p'_j=p_j} (t(j)-p(j)). \label{haaronleed}
\end{align}
Note that the inner two sums above define a quadratic form. The singular values are equal to the absolute value of the eigenvalues of a symmetric matrix and the largest singular value of a matrix is upper bounded by the Frobenius norm of the Hessian. Then from basic theory of quadratic form optimization \cite[7.2]{davidlay1}, we have
\begin{align*}
    &\Bigg| \sum_{i=1}^J \sum_{j=1}^J  (t(i)-p(i)) \frac{\partial^2 I_l(Q^{p,d,\rho},p',d,\rho)}{\partial p'_i \partial p'_j} \Bigg |_{p'_i = p_i, p'_j = p_j}  (t(j)-p(j)) \Bigg |\\
    &\,\,\,\,\,\,\,\,\,\,\,\,\,\,\,\,\,\,\,\,\,\,\,\,\,\,\,\,\,\,\,\,\,\,\,\,\,\,\,\,\,\,\leq \Bigg | \Bigg | \,\,\frac{\partial^2 I_l(Q^{p,d,\rho},p',d,\rho)}{\partial p'^2} \Bigg |_{p' = p} \,\,\Bigg | \Bigg |_F \cdot \sum_{j=1}^J (t(j) - p(j))^2.
\end{align*}
Hence, the absolute value of  $(\ref{haaronleed})$ is upper bounded by
\begin{align}
    &\frac{1}{2} \Bigg | \Bigg | \,\,\frac{\partial^2 I_l(Q^{p,d,\rho},p',d,\rho)}{\partial p'^2} \Bigg |_{p' = p} \,\, \Bigg | \Bigg |_F \cdot \sum_{j=1}^J \sum_{t \in \mathcal{P}_n(A)} p^n(T^n_A(t)) (t(j) - p(j))^2 \notag \\
    &= \frac{1}{2} \Bigg | \Bigg | \,\,\frac{\partial^2 I_l(Q^{p,d,\rho},p',d,\rho)}{\partial p'^2} \Bigg |_{p' = p} \,\, \Bigg | \Bigg |_F \cdot \sum_{j=1}^J \mathbb{E}_p \left [ (T(j) - p(j))^2 \right ] \notag \\
    &= \frac{1}{2} \Bigg | \Bigg | \,\,\frac{\partial^2 I_l(Q^{p,d,\rho},p',d,\rho)}{\partial p'^2} \Bigg |_{p' = p} \,\, \Bigg | \Bigg |_F \cdot  \sum_{j=1}^J \frac{p(j) (1-p(j))}{n} \notag \\
    &= O\left(\frac{1}{n} \right).  \label{resulta2n}
\end{align}
Hence, substituting $(\ref{resulta1niya})$ and $(\ref{resulta2n})$ into $(\ref{subeeresult2})$ gives 
\begin{align*}
    \mathbb{E}_p \left [ R(T,d,\rho) \right ] &\leq R(p,d,\rho) + o\left(\frac{\ln n}{n} \right).
\end{align*}

\end{IEEEproof}


\section{Proof of Lemma \ref{dballlemma} \label{dballlemmaproof}}

Fix any $d > 0$ and $(p, \rho) \in \mathcal{S}_d$. Let $\delta > 0$ be a number to be specified later. Let $(t,\rho') \in \mathcal{N}_\delta(p, \rho)$ and $x^n \in T^n_A(t)$ be any sequence within the type class. By the Definition of $\mathcal{S}_d$, there exists a $\sigma > 0$ such that 
\begin{itemize}
    \item $p(j) \geq \sigma$ for all $j \in A$,
    \item $Q^{p,d,\rho}(k) \geq \sigma$ for all $k \in B$, and 
    \item $0 < d < \min_{k \in B} \sum_{j \in A} t(j) \rho(j, k)$. 
\end{itemize}  
The last condition above implies that we must have that $\rho(j, k) > d$ for some $j \in A$ and $k \in B$. From the definition of $\mathcal{S}_d$ in Definition \ref{defineSdset} and the continuity of $Q^{p,d,\rho}$ in $p$ and $\rho$ which is implied by Lemma \ref{contlemma},    it is easy to see that we can make $\delta$ small enough such that 
\begin{itemize}
    \item $t(j) \geq \sigma/2$ for all $j \in A$,
    \item $Q^{t,d,\rho'}(k) \geq \sigma/2$ for all $k \in B$, and 
    \item $0 < d < \min_{k \in B} \sum_{j \in A} t(j) \rho'(j, k)$, 
\end{itemize}  
for all $(t,\rho') \in \mathcal{N}_\delta(p,\rho)$. The last condition above also trivially implies that $\rho'(j, k) > d$ for some $j \in A$ and $k \in B$.

For the given sequence $x^n \in T^n_A(t)$ and distortion measure $\rho'$, define a sequence of independent random variables $U_1, U_2,$ $\ldots, U_n $ as  
\begin{align}
    U_i &\triangleq \rho'(x_i, \tilde{Y}_i) - \sum_{k \in B} W^*_{B|A}(k|x_i) \rho'(x_i, k), \label{loopaUi's}
\end{align}
where we write $W^*_{B|A} = W^*_{B|A}[t,d,\rho']$ and $\tilde{Y}_i \sim W^*_{B|A}(\cdot | x_i)$. Clearly, each $U_i$ has finite second- and third-order moments which we denote by $\mathbb{E}[U_i^2] = \nu_i^2 $ and $\mathbb{E}[|U_i|^3] = \eta_i$. We have that 
\begin{align}
    \sum_{i=1}^n \eta_i \leq n \rho_{\max}^3.
\end{align}
Next, we show that $\sum_{i=1}^n \nu_i^2$ also grows linearly with $n$. 

\begin{fact}
From Equation $(10.124)$ in \cite{thomas_cov}, we have the following relation between $W^*_{B|A}[t,d,\rho']$ and $Q^{t,d,\rho'}$:
\begin{align*}
    W^*_{B|A}[t,d,\rho'](k|j) = \frac{Q^{t,d,\rho'}(k) \exp \left(- \lambda^* \rho'(j,k)  \right)}{\sum \limits_{k' \in B} Q^{t,d,\rho'}(k') \exp \left( -\lambda^* \rho'(j,k') \right) },
\end{align*}
where $-\lambda^* = \partial R(t,d,\rho') / \partial d$.  \label{jimmyneutron}
\end{fact}
Hence, it follows that  support$(Q^{t,d,\rho'}) = K$ if and only if $W^*_{B|A}[t,d,\rho'](k |j) > 0$ for all $j \in A$ and $k \in B$. In fact, since $Q^{t,d,\rho'}(k) \geq \sigma/2$ for all $k$, we have 
\begin{align*}
    W^*_{B|A}[t,d,\rho'](k|j) &\geq \frac{\sigma}{2} \exp \left(- \lambda^* \rho_{\max} \right)\\
    &\geq \frac{\sigma}{2} \exp \left(- \frac{\rho_{\max}}{d} \ln (K)  \right)
\end{align*}
for all $j \in A$, $k \in B$, where the last inequality above follows by using the assumption in $(\ref{dist_assump})$, which implies that (i) $R(t,d,\rho') \leq \ln(K)$ and (ii) $ \lambda^*=  | \partial R(t,d,\rho')/\partial d | \leq \ln(K)/ d$ by convexity of $R(t,d,\rho')$ in $d$.

Since we have (i) a zero in every row of the distortion matrix $\rho'$, (ii) $\rho'(j^*, k) > d$ for some $j^* \in A$ and $k \in B$, (iii) $W^*_{B|A}[t,d,\rho'](k|j) > \frac{\sigma}{2} \exp\left( - \rho_{\max} \ln (K)/d \right)$ for all $k$ and $j$, and (iv) $t(j) \geq \sigma/2$ for all $j \in A$,  we have 
\begin{align*}
    \sum_{i=1}^n \nu_i^2 \geq \frac{n \sigma}{2} \nu_{i^*}^2 > 0. 
\end{align*}
where $i^*$ satisfies $x_{i^*} = j^*$. Consider the $j^*$th row of the distortion matrix whose entries include $0$ and $d'$, where $d' > d > 0$. There is a full-support distribution $W^*_{B|A}[t,d,\rho'](\cdot | j^*)$ over the entries of this row  and each entry of the distribution is uniformly bounded away from zero in terms of $\sigma, \rho_{\max}$ and $d$. Hence, the variance of the random variable $U_{i^*}$ can be uniformly bounded away from zero by a number which depends only on $\sigma, \rho_{\max}$ and $d$. Hence, we can write 
\begin{align*}
    \sum_{i=1}^n \nu_i^2 \geq n \,\nu_{\min}^2 > 0. 
\end{align*}
We now invoke the Refined Lucky-Strike Lemma \cite[Lemma 8]{https://doi.org/10.48550/arxiv.2202.04481} which, specialized to the $\epsilon = 0$ case, establishes that for any positive number $C$, 
\begin{align}
    &\mathbb{P}\left (\rho'_n(x^n, Y^n) \leq d\right)\notag\\ 
    &\geq \exp \left(-n R(t,d,\rho') - C \lambda^* \right)   \mathbb{P} \left( - C \leq \sum_{i=1}^n U_i \leq 0 \right) \notag \\
    &\geq \exp \left(-n R(t,d,\rho') - C \ln K / d \right)   \mathbb{P} \left( - C \leq \sum_{i=1}^n U_i \leq 0 \right) \label{further14}
\end{align}
for all integers $n$ and $x^n \in T^n_A(t)$, where $Y^n$ is distributed according to $(Q^{t,d,\rho'})^n$, where $\lambda^* = -\partial R(t,d,\rho') / \partial d $ and $\lambda^* \leq \ln(K) / d$ by the same argument as before. 

We continue $(\ref{further14})$ as 
\begin{align}
    &\mathbb{P}\left (\rho'_n(x^n, Y^n) \leq d\right)\notag\\ 
    &\geq \exp \left(-n R(t,d,\rho') - C \ln (K) / d \right)   \mathbb{P} \left( - \frac{C}{\sqrt{\sum_{i=1}^n \nu_i^2}} \leq \frac{\sum_{i=1}^n U_i}{\sqrt{\sum_{i=1}^n \nu_i^2}} \leq 0 \right) \notag \\
    &\geq \exp \left(-n R(t,d,\rho') - C \ln(K)/d \right)   \left [ F_n(0) - F_n \left(- \frac{C}{\sqrt{\sum_{i=1}^n \nu_i^2}}\right) \right ], \label{leftcodm}
\end{align}
where $F_n$ denotes the cumulative distribution function of $\frac{\sum_{i=1}^n U_i}{\sqrt{\sum_{i=1}^n \nu_{i}^2}}$. Now by Berry-Esseen theorem \cite{esseen11}, we have that for all $n$ there exists an absolute constant $C_0$ such that 
\begin{align}
    \sup_{s \in \mathbb{R}} |F_n(s) - \Phi(s)| \leq C_0 \left( \sum_{i=1}^n \nu_i^2 \right)^{-3/2} \sum_{i=1}^n \eta_i. \label{Ihategaming}
\end{align}
Using the bounds for the second- and third-order moments in the preceding discussion, we have 
\begin{align}
    \sup_{s \in \mathbb{R}} |F_n(s) - \Phi(s)| \leq \frac{C_0}{\sqrt{n}}   \frac{\rho_{\max}^3}{\nu_{\min}^3}. \label{realizationofberry}
\end{align}

Continuing $(\ref{leftcodm})$ using $(\ref{realizationofberry})$, we have 
\begin{align}
    &\mathbb{P}\left (\rho'_n(x^n, Y^n) \leq d\right)\notag\\ 
    &\geq \exp \left(-n R(t,d,\rho') - C \ln(K)/d \right)   \left [ \frac{1}{2} - \Phi \left(- \frac{C}{\sqrt{\sum_{i=1}^n \nu_i^2}}\right) - 2 \frac{C_0}{\sqrt{n}}   \frac{\rho_{\max}^3}{\nu_{\min}^3} \right ] \label{nidaaleena}
\end{align}
For the first two terms inside the brackets in $(\ref{nidaaleena})$, we have the following lower bound: 
\begin{align}
    &\frac{1}{2} - \Phi \left(- \frac{C}{\sqrt{\sum_{i=1}^n \sigma_i^2}}\right) \notag \\
    &\geq \frac{1}{2} - \Phi \left(- \frac{C}{\rho_{\max} \sqrt{n}}\right) \notag \\
    &= \frac{1}{\sqrt{2 \pi}}\int_{- \frac{C}{\rho_{\max} \sqrt{n}}}^{0} e^{-x^2/2} dx \notag  \\
    &\geq \frac{1}{\sqrt{2 \pi}}\int_{- \frac{C}{\rho_{\max} \sqrt{n}}}^{0} \left( 1 - \frac{x^2}{2}\right) dx \notag \\
    &= \frac{1}{\sqrt{2 \pi}}  \frac{C}{\rho_{\max} \sqrt{n}} - \frac{1}{6 \sqrt{2\pi}} \frac{C^3}{\rho_{\max}^3 n^{3/2}}. \label{nonminimaxlowerbnd}
\end{align}
Finally, using $(\ref{nonminimaxlowerbnd})$ back in $(\ref{nidaaleena})$, we obtain 
\begin{align}
    &\mathbb{P}\left (\rho'_n(x^n, Y^n) \leq d\right)\notag\\
    &\geq \exp \left(-n R(t,d,\rho') - C \ln(K)/d \right) \left [\frac{1}{\sqrt{2 \pi}}  \frac{C}{\rho_{\max} \sqrt{n}} - \frac{1}{6 \sqrt{2\pi}} \frac{C^3}{\rho_{\max}^3 n^{3/2}} - 2 \frac{C_0}{\sqrt{n}}   \frac{\rho_{\max}^3}{\nu_{\min}^3}   \right ] \notag \\
    &\stackrel{(a)}{\geq} \exp \left(-n R(t,d,\rho') - C \ln(K)/d \right) \left [\frac{1}{2\sqrt{2 \pi}}  \frac{C}{\rho_{\max} \sqrt{n}}  - 2 \frac{C_0}{\sqrt{n}}   \frac{\rho_{\max}^3}{\nu_{\min}^3}   \right ] \notag \\
    &\stackrel{(b)}{=} \exp \left(-n R(t,d,\rho') - \frac{1}{2} \ln n  + O(1)\right). \label{lastranopqe}
\end{align}
where inequality $(a)$ follows by assuming $C^2 \leq 3 \rho_{\max}^2 n$, and equality $(b)$ follows by allowing sufficiently large $n$ to allow the choice of the free parameter $C$ to satisfy  
$$\frac{1}{2\sqrt{2 \pi}}  \frac{C}{\rho_{\max} }  - 2 C_0   \frac{\rho_{\max}^3}{\nu_{\min}^3} > 0,$$
where one can use the upper bound $C_0 \leq 0.56$ \cite{Shevtsova}. The $O(1)$ only depends on $p$, $d$,  $\rho$ and $\rho_{\max}$. We omit the dependence on $\sigma$ or $\delta$ because $\sigma$ and $\delta$ themselves depend on the aforementioned variables.

\section{Proof of Lemma \ref{lemmagolden} \label{lemmagoldenproof}}

Fix $d > 0$ and let $\rho$ be the input distortion measure. Define 
\begin{align*}
    B(y^n, t,d,\rho) \triangleq \{ x^n \in T^n_A(t) : \rho_n(x^n, y^n) \leq d \}
\end{align*}
to be the set of type $t$ source sequences covered within distortion $d$ by a reconstruction sequence $y^n$. The distortion constraint $\rho_n(x^n, y^n) \leq d$ can be written in terms of types; denoting the type of $y^n$ by $t_y$ and the conditional type of $x^n$ given $y^n$ by $W$, we have 
\begin{align}
    \rho_n(x^n, y^n) &= \frac{1}{n} \sum_{i=1}^n \rho(x_i, y_i) \notag \\
    &= \sum_{j \in A, k \in B} t_y(k) W(j|k) \rho(j, k) \leq d. \label{heenaamaya}
\end{align}
Let 
\begin{align*}
    \mathcal{C}(y^n, t,d,\rho) \triangleq &\left \{ W  : \sum_k t_y(k) W(j|k) = t(j) \,\,\,\forall j \in A \right. \\
    &\left . \text{and} \,\,\,\,\,\,\sum_{j,k} t_y(k) W(j|k) \rho(j, k) \leq d  \right \}    
\end{align*}
be the set of all conditional types satisfying the given constraints. For a fixed $y^n$, the number of conditional types of $x^n$ given $y^n$ is at most $(n+1)^{JK - 1}$; hence, $|\mathcal{C}(y^n, t,d,\rho)| \leq (n+1)^{JK - 1}$. The size of $B(y^n, t,d,\rho)$ can then be evaluated by summing the sizes of the conditional type classes $T_W(y^n)$ of all the conditional types $W \in \mathcal{C}(y^n, t, d, \rho)$. From \cite[Lemma 2.3 and Lemma 2.5]{korner1}, we have the following bounds for $|T^n_A(t)|$ and $|T_W(y^n)|$:\footnote{In the cited 
    reference, the lower bounds are stated with powers $J$ and $JK$
    instead of $J-1$ and $JK-1$, respectively, but the bounds as
   stated here evidently hold as well.}
\begin{align}
    &\frac{1}{(n+1)^{J-1}} \exp \left( n H(t) \right) \leq |T^n_A(t)| \leq \exp \left( nH(t)\right) \text{ and} \label{boundtypecond330}\\
    &\frac{1}{(n+1)^{JK-1}} \exp \left( n H(W|t_y) \right) \leq |T_W(y^n)| \notag \\
    &\,\,\,\,\,\,\,\,\,\,\,\,\,\,\,\,\,\,\,\,\,\,\,\,\,\,\,\,\,\,\,\,\,\,\,\,\,\,\,\,\,\,\,\,\,\,\,\,\,\,\,\,\,\,\,\,\,\,\,\,\,\,\,\,\,\,\,\,\,\leq \exp \left( nH(W|t_y)\right). \label{boundtypecond}
\end{align}
Equipped with these, we evaluate the size of $B(y^n, t,d,\rho)$ as follows:
\begin{align}
    &|B(y^n, t,d,\rho)| \notag\\
    &= \sum_{W \in \mathcal{C}(y^n, t, d, \rho)} |T_W(y^n)| \notag \\
    &\leq \sum_{W \in \mathcal{C}(y^n, t, d, \rho)} e^{n H(W|t_y) }  \notag  \\
    &\leq (n+1)^{JK - 1} \exp \left( n \max_{W \in \mathcal{C}(y^n, t,d,\rho)  } H(W|t_y) \right). \label{ballupperbound}
\end{align}
Now let $X^n \sim p^n$ and let $Y^n = g_n(f_n(X^n, \rho))$. We then have 
\begin{align}
    &H(Y^n | t(X^n) = t) \notag \\
    &= -\sum_{y^n \in B^n} \mathbb{P}\left(Y^n = y^n|t(X^n) = t \right) \ln \left( \mathbb{P}\left(Y^n = y^n|t(X^n) = t \right) \right) \notag \\
    &= -\sum_{y^n \in B^n} \mathbb{P}\left(Y^n = y^n|t(X^n) = t \right) \ln \left( \frac{\mathbb{P}\left(Y^n = y^n, t(X^n) = t \right)}{\mathbb{P}(t(X^n) = t)} \right) \notag \\
    &\geq -\sum_{y^n \in B^n} \mathbb{P}\left(Y^n = y^n|t(X^n) = t \right) \ln \left( \frac{|B(y^n, t,d,\rho)|}{|T^n_A(t)|} \right) \notag \\
    &\geq -\sum_{y^n \in B^n} \mathbb{P}\left(Y^n = y^n|t(X^n) = t \right) \left [ \ln (n+1)^{JK+J-2} - \mbox{ }\right. \notag \\
    &\,\,\,\,\,\,\,\,\,\,\,\,\,\,\,\,\,\,\,\,\,\, \left . n \left( H(t) - \max_{W \in \mathcal{C}(y^n, t,d,\rho)} H(W|t_y) \right)  \right], \label{expandmaxrd}
\end{align}
where the last inequality above uses $(\ref{boundtypecond330})$ and $(\ref{ballupperbound})$. To continue $(\ref{expandmaxrd})$, we note that $H(W|t_y)$ is a function of the joint distribution, call it $s \in \mathcal{P}(A \times B)$, specified by $t_y$ and $W$. Let 
\begin{align*}
    \mathcal{C}^*(t,d,\rho) \triangleq &\left \{ s \in \mathcal{P}(A \times B) : \sum_k s(j,k) = t(j) \,\,\,\forall j \in A \right .\\
    &\,\,\,\,\,\,\,\,\,\,\,\,\,\,\,\left  . \text{and}\,\,\, \,\,\,\,\,\,\sum_{j,k} s(j,k) \rho(j, k) \leq d  \right \}.
\end{align*}
It is easy to see that if $(\tilde{X}, \tilde{Y}) \sim s$, then  
$$\max_{W \in \mathcal{C}(y^n, t,d,\rho)} H(W|t_y) \leq \max_{s \in \mathcal{C}^*(t,d,\rho)} H(\tilde{X}|\tilde{Y}).$$
Then, using the definition of the rate-distortion function, we can continue $(\ref{expandmaxrd})$ as 
\begin{align*}
    &H(Y^n|t(X^n) = t)\\
    &\geq \sum_{y^n \in B^n} \mathbb{P}\left(Y^n = y^n|t(X^n) = t \right) \left [ n R(t,d,\rho) -\ln (n+1)^{JK+J-2}   \right]\\
    &= n R(t,d,\rho) - \ln (n+1)^{JK+J-2}.
\end{align*}
To finish the proof, we use the fact that for any prefix code, the expected length is lower bounded by the entropy. Hence,  
\begin{align}
    &\frac{1}{n} \mathbb{E}_p \left [ l(f_n(X^n, \rho)) \right ] \notag \\
    &\geq \frac{1}{n} H(Y^n) \notag \\
    &\geq \frac{1}{n} H(Y^n|t(X^n)) \notag \\
    &= \frac{1}{n} \sum_{t \in \mathcal{P}_n(A)} p^n(T^n_A(t)) H(Y^n | t(X^n) = t) \notag \\
    &\geq \frac{1}{n} \sum_{t \in \mathcal{P}_n(A)} p^n(T^n_A(t)) \left ( n R(t,d,\rho) - \ln (n+1)^{JK+J-2} \right) \notag \\
    &\geq \mathbb{E}_p \left [ R(T,d,\rho) \right] - (JK + J - 2)\frac{\ln n }{n} - \frac{JK + J - 2}{n}. \notag
\end{align}

\section*{Acknowledgment}

The authors would like to thank En-hui Yang for supplying a copy of his unpublished work \cite{yang3} and for helpful discussions. This research was supported by the US National Science Foundation
under grants CCF-2008266, CCF-1934985 and CCF-1956192, by the US Army Research
Office under grant W911NF-18-1-0426 and by a gift from Google.


\ifCLASSOPTIONcaptionsoff
  \newpage
\fi



\bibliographystyle{IEEEtran}
\end{document}